\documentclass[12pt]{report}

\usepackage{setspace}
\usepackage{fullpage}
\usepackage{amsmath}
\usepackage{graphicx}
\usepackage{titlesec}

\usepackage{slashed}
\usepackage{amssymb}
\usepackage{amsfonts}
\usepackage{geometry}

\def\bea{\begin{eqnarray}}
\def\eea{\end{eqnarray}}
\def\nn{\nonumber}
\def\lb{\Lambda_b}

\def\lq{\textquotedblleft}
\def\rq{\textquotedblright}
\def\lsq{\textquoteleft}
\def\rsq{\textquoteright}

\def\lblc{\Lambda_{b}\to\Lambda_{c}\tau\bar{\nu}_{\tau}}
\def\bcl{\Lambda_{b}\to\Lambda_{c} \ell \bar{\nu}_{\ell}}
\def\bc{b\to c l^{-} \bar{\nu}_l}

\titleformat{\chapter}[display]% NEW
    {\normalsize\centering}{\MakeUppercase{\chaptertitlename}\ \thechapter}{0pt}{\normalsize}% NEW
\titleformat*{\subsection}{\normalsize}

\titlespacing*{\chapter}{0pt}{45pt}{12pt}% NEW
\titlespacing*{\section}{0pt}{24pt}{12pt}% NEW
\titlespacing*{\subsection}{0pt}{24pt}{0pt}% NEW

\begin{document}

\newgeometry{left=1in, right=1in}
\bibliographystyle{Bibfiles/agu04}

%%%%%%%%%%%%%%%%%%%%%%%%%%%%%%%%
%title page---------------------
\setlength{\oddsidemargin}{0in}
\setlength{\textwidth}{6.5in}
\setlength{\topmargin}{1in}
\setlength{\textheight}{9in}
\newpage
\pagenumbering{roman}

\begin{titlepage}
\begin{center}
	\begin{Large}
	SEMI-LEPTONIC DECAY OF LAMBDA-B 	
	\singlespacing	
	IN THE STANDARD MODEL AND WITH NEW PHYSICS
	\end{Large}
	\vspace{1.8in}
	\singlespacing
	A Thesis\par
	\singlespacing
	presented in partial fulfillment of requirements\par
	\singlespacing	
	for the degree of Master of Science\par
	\singlespacing
	in the Department of Physics and Astronomy\par
	\singlespacing
	The University of Mississippi \par
	\vspace{1.8in}
	\doublespacing
	by\par	
	WANWEI WU\par
	April 2015
\end{center}
\end{titlepage}

%%%%%%%%%%%%%%%%%%%%%%%%%%%%%%%%%%%%%%%%%%%
%copy right--------------------------------
\clearpage
\setlength{\oddsidemargin}{0in}
\setlength{\textwidth}{6.5in}
\setlength{\topmargin}{0in}
\setlength{\textheight}{9in}
\thispagestyle{empty}
\doublespacing
\vspace*{\fill}
\begin{center}
Copyright \copyright~2015 by Wanwei Wu\\
ALL RIGHTS RESERVED.
\end{center}

%%%%%%%%%%%%%%%%%%%%%%%%%%%%%%%%%%%%%%%%%%%%%%%%
%abstract---------------------------------------
\newpage
\setlength{\oddsidemargin}{0in}
\setlength{\textwidth}{6.5in}
\setlength{\topmargin}{0in}
\setlength{\textheight}{9in}
\doublespacing
\chapter*{\centerline{ABSTRACT}}
\addcontentsline{toc}{chapter}{ABSTRACT}

\hspace*{\parindent}Heavy quark decays provide a very advantageous investigation to test the Standard Model (SM). Recently, promising experiments with \textit{b} quark, as well as the analysis of the huge data sets produced at the B factories, have led to an increasing study and sensitive measurements of relative \textit{b} quark decays. In this thesis, I calculate various observables in the semi-leptonic decay process $\lblc$ both in the SM and in the presence of New Physics (NP) operators with different Lorentz structures. The results are relevant for the coming measurement of this semi-leptonic decay at LHC \textit{b} experiment in CERN, and also provide theoretical predictions to refine the physics beyond the SM.

%%%%%%%%%%%%%%%%%%%%%%%%%%
%%%%%%%%%%%%%%%%%%%%%%%%%%%%%%%%%%%%%%%%%%%%%%%%%%%%%%%%%%%
%acknowledgements------------------------------------------
\newpage
\setlength{\oddsidemargin}{0in}
\setlength{\textwidth}{6.5in}
\setlength{\topmargin}{0in}
\setlength{\textheight}{9in}
\doublespacing
\chapter*{\centerline{ACKNOWLEDGEMENTS}}
\addcontentsline{toc}{chapter}{ACKNOWLEDGEMENTS}

\hspace*{\parindent}I would like to express my sincere gratitude to my advisor Dr. Alakabha Datta for his continuous support. Without his guidance, I could not have finished this thesis. His valuable advice and comments, as well as his patience and immense knowledge, helped me in all the time with my study and research. I would like to thank the rest of my thesis committees: Dr.\ Lucien Cremaldi and Dr.\ Luca Bombelli for their insightful comments and precious time.

Also, my sincere thanks go to Dr.\ Emanuele Berti, Dr.\ Donald Summers, Dr.\ Breese Quinn, Dr.\ Murugeswaran Duraisamy, Dr.\ Preet Sharma, and Shanmuka Shivashankara. In particular, I would like to thank Hongkai Liu for studying together and helpful discussions.  

In addition, I would like to thank my parents. As simple and kind-hearted farmers, they try to understand and support me all the way.

This work was financially supported in part by the National Science Foundation under Grant No.NSF PHY-1414345.
\clearpage

%%%%%%%%%%%%%%%%%%%%%%%%%%%%%%%%%%%%%%%%%%%%%%%%%%%%%%%%%%%%%
%content-----------------------------------------------------
\newpage
\doublespacing
\setlength{\oddsidemargin}{0in}
\setlength{\textwidth}{6.5in}
\setlength{\topmargin}{0in}
\setlength{\textheight}{9in}
\tableofcontents

%%%%%%%%%%%%%%%%%%%%%%%%%%%%%%%%%%%%%%%%%%%%%%%%%%%%%%%%%%%%%
%list of tables----------------------------------------------
\newpage
\label{listoftable}
\addcontentsline{toc}{chapter}{LIST OF TABLES}
\clearpage

\listoftables

%%%%%%%%%%%%%%%%%%%%%%%%%%%%%%%%%%%%%%%%%%%%%%%%%%%%%%%%%%%%%
%list of figures----------------------------------------------
\newpage
\label{listoffig}
\addcontentsline{toc}{chapter}{LIST OF FIGURES}
\listoffigures
\clearpage

%%%%%%%%%%%%%%%%%%%%%%%%%%%%%%%%%%%%%%%%%%%%%%%%%%%%%%%%%%%%%%%
%chapter 1-----------------------------------------------------

\newpage
\pagenumbering{arabic}

\doublespacing
\titleformat{\section}{\normalsize}{\thesection}{1em}{}
\newpage
\pagenumbering{arabic}

\chapter{INTRODUCTION}

\singlespacing
\hspace*{\parindent}For the past several decades, the Standard Model (SM) has been the most successful theory concerning the fundamental particles and most of their interactions, namely the electromagnetic, weak and strong forces. It has not only successfully explained almost all the elementary particle experiment results so far, but precisely predicted a very wide variety of phenomena, leading us to a better understanding of the fundamental structure of matter. Specially, the discovery of the Higgs boson \cite{atlas,cms}, which is a scalar particle,  makes the SM a remarkably successful description of the subatomic world.  

However, there are some things the SM still cannot explain, i.e., the mass of neutrino, the dark matter and the dark energy, and even the most familiar force in our everyday life--gravity. Therefore, finding physics beyond the SM becomes a major part of particle physics. In this explorative process of New Physics (NP), both the third generation charged leptons and the third generation quarks play important roles since they are comparatively heavier and also relatively more sensitive to NP. In addition, the constraints on NP involving the third generation leptons ($\tau$ and $\nu_{\tau}$ ) and quarks ($b$ and $t$) are somewhat weaker, leading to possible larger NP effects.

Heavy quark decays provide a very advantageous investigation to test the SM. Recently, the BaBar Collaboration has reported their measurements of the ratio of the branching fractions of $\bar{B}\to D^{(*)} \tau^{-} \bar{\nu}_{\tau}$ to $\bar{B}\to D^{(*)} \ell^{-} \bar{\nu}_{\ell}$ \cite{BaBar1,BaBar2}:
\begin{eqnarray}
R(D)&\equiv& \frac{{\cal B}(\bar{B} \to D^{+}\tau^{-} \bar{\nu}_{\tau})}{{\cal B}( \bar{B} \to D^{+}\ell^{-} \bar{\nu}_{\ell})}=0.440 \pm 0.058 \pm 0.042, \nonumber \\R(D^{*})&\equiv& \frac{{\cal B}(\bar{B}\to D^{*+}\tau^{-}\bar{\nu}_{\tau})}{{\cal B}(\bar{B}\to D^{*+}\ell^{-} \bar{\nu}_{\ell})}=0.332 \pm 0.024 \pm 0.018,
\label{babar}
\end{eqnarray}
where $\ell = e,\mu$. However, the SM predictions for $R(D)$ and $R(D^{*})$ are \cite{BaBar1,BDSM1,BDSM2}
\begin{eqnarray}
R(D)&=& 0.297 \pm 0.017, \nonumber \\
R(D^{*})&=& 0.252 \pm 0.003,
\label{bdsm}
\end{eqnarray}
which deviate from the BaBar measurements by $2\sigma$ and $2.7\sigma$, respectively. (The BaBar Collaboration itself reported a 3.4$\sigma$
deviation from SM when the two measurements of Eq.~(\ref{babar})
are taken together.) These non-universality deviations could be providing a hint of NP \cite{bdnew1, bdnew2, bdnew3, bdnew4}. Another possible test of such a non-universality can be in the semi-leptonic $\lblc$ decay, which has not been measured experimentally though it might be measured at LHC b experiment in CERN soon. In both $\bar{B}$ meson and $\Lambda_{b}$ baryon decays, the underlying quark level transition $b\to c\tau^{-}\bar{\nu}_{\tau}$ can be probed, as both $\bar{B}$ meson and $\Lambda_{b}$ baryon contain a $b$ quark which will decay here.

In this thesis, I calculate various observables in the semi-leptonic decay process $\lblc$ both in the SM and in the presence of NP operators with different Lorentz structures by using constraints on the NP couplings obtained by using Eq.~(\ref{babar}). Since the calculations involve the structures of both $\bar{B}$ meson and $\Lambda_{b}$ baryon, the Quantum Chromodynamics (QCD) for the strong interactions between quarks and gluons (specially, the form factors), will be briefly introduced as well as the SM and the weak interactions.

%%%%%%%%%%%%%%%%%%%%%%%%%%%%%%%%%%%%%%%%%%%%%%%%%%%%%%%%%%%%%%%%%%%%%%
%standard model-------------------------------------------------------
\section{Standard Model}

\begin{figure}
\begin{center}
\includegraphics[width=15cm, height=8.5cm]{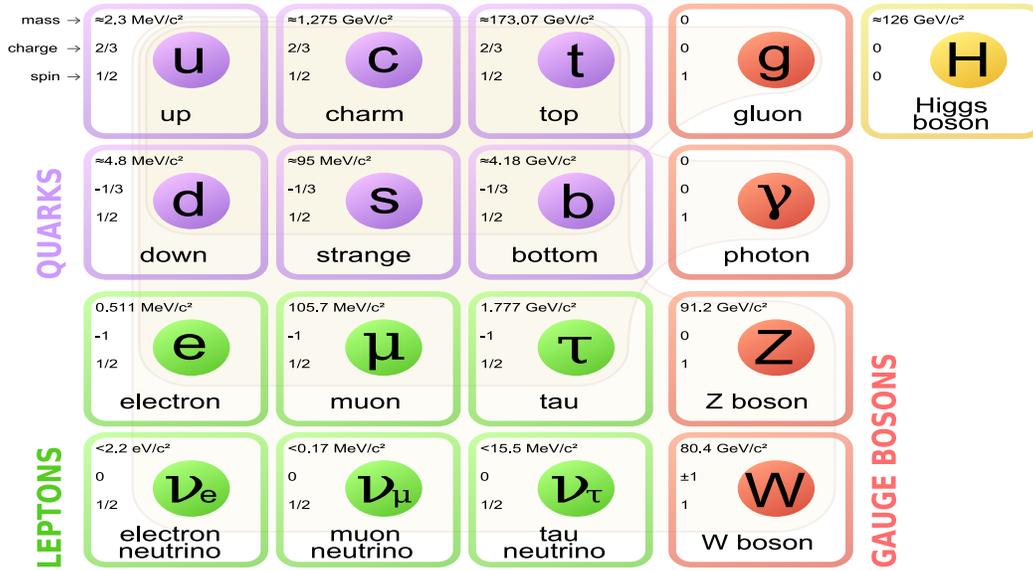}
\end{center}
\caption{The SM of Elementary Particles (matter fermions in the first three generations, gauge bosons in the fourth column, and the Higgs boson in the fifth)}
\label{FSM}
\end{figure}

\hspace*{\parindent}The Standard Model of particle physics, formulated in the 1970s, is a theory of fundamental particles and their interactions. It is based on the quantum theory of fields and provides the most accurate description of nature at the subatomic level so far. According to this model, all matter is built from a small number of fundamental spin-$\frac{1}{2}$ particles, called \textit{fermions}: six \textit{quarks} and six \textit{leptons}, which follow the Fermi-Dirac statistics; while the carriers of the interactions are characterized as \textit{bosons}, which possess integer spin (either 0 or 1 ) and follow the Bose-Einstein statistics. There are seventeen named particles in the SM, which are organized in Fig.~\ref{FSM}. The Higgs boson, as the last particle in the SM, was discovered in 2012 \cite{atlas,cms}.

There are four known fundamental interactions in the universe: the gravitational, the electromagnetic, the weak and the strong interactions. They work over different ranges and have different strengths. Gravity, acting between all types of particle, is the weakest but it has an infinite range. It is supposedly mediated by exchange of a spin-2 boson, the \textit{graviton}, which has not been observed. Even though it is universal and is dominant on the scale of the universe, gravity is not included in the SM because it is much weaker than the other forces  and can be neglected at the level of individual subatomic particles. The electromagnetic interaction acts between all charged particles and is mediated by \textit{photon} ($\gamma$) exchange. It also has infinite range but it is many times stronger than gravity. The weak and strong interactions are effective only over a very short range and dominate only at the level of subatomic particles. The weak interaction is associated with the exchange of elementary spin-1 bosons between quarks and/or leptons. These mediators are $W^{\pm}$ and $Z^{0}$ bosons, with masses of order 100 times the proton mass. The strong interaction, as its name suggests, is the strongest of all four fundamental interactions. It is responsible for binding the quarks in the neutron and proton, and the neutrons and protons within nuclei. The strong force is mediated by spin-1, massless particles known as \textit{gluons}, which couple to color charge, rather like the photons couple to electromagnetic charge.

Fermions are fundamental matter particles in the SM. These twelve particles (six \textit{leptons} and six \textit{quarks}) can be grouped into three generations. The lightest and most stable particles make up the first generation, whereas the heavier and less stable particles belong to the second and third generations. The \textit{leptons} carry integral electric charge. The charged \textit{leptons} are the electron, muon and tau, while the neutral leptons are the corresponding neutrinos. A different \textquotedblleft flavour\textquotedblright \ of neutrino is paired with each \lq flavour\rq \ of charged lepton, as indicated by the subscript, i.e.,  ($e$, $\nu_{e}$), ($\mu$, $\nu_{\mu}$) and ($\tau$, $\nu_{\tau}$). The charged muon and tau are both unstable and decay spontaneously to electrons, neutrinos and other particles. The mean lifetime of the muon is $2.2\times 10^{-6}$ s, that of the tau only $2.9\times 10^{-13}$ s. Neutrinos were postulated by Pauli in 1930 in order to account for the energy and momentum missing in the process of nuclear $\beta$-decay. They experience the weak interactions only. The \textit{quarks} carry fractional electric charges, of $+\frac{2}{3}e$ or $-\frac{1}{3}e$. The quark \lq flavour\rq\ is denoted by a symbol: $u$ for \lsq up\rsq, $d$ for \lsq down\rsq, $s$ for \lsq strange\rsq, $c$ for \lsq charmed\rsq, $b$ for \lsq bottom\rsq\ and $t$ for \lsq top\rsq. While leptons exist as free particles, quarks are not found to do so. The bound states of quarks are called \textit{hadrons}, which can be categorized into two families: baryons (made of three quarks) and mesons (made of one quark and one anti-quark). Each quark carries one of the three colors(or color charges): $r$, $g$ and $b$. Quarks are bound together by \textit{gluons}, which are also colored. Fig.\ref{FSM} shows that the three lepton pairs are exactly matched by the three quark pairs.

%%%%%%%%%%%%%%%%%%%%%%%%%%%%%%%%%%%%%%%%%%%%%%%%%%%%%%%%%%%%%%%%%%%%%%
%weak interaction-------------------------------------------------------
\section{Weak Interactions}

\hspace*{\parindent} The weak interaction is mediated by three massive bosons, the charged $W^{\pm}$ and the neutral $Z^{0}$. The $W^{+}$ and $W^{-}$ are anti-particles of each other, while the $Z^{0}$, like the photon, is its own anti-particle. Depending on whether leptons and/or hadrons are involved, the weak interaction can be conventionally divided into three categories: (i) purely leptonic processes, e.g., $\mu^{-}\to e^{-}+\bar{\nu}_{e}+\nu_{\mu}$, (ii)  semi-leptonic processes involving both hadrons and leptons, e.g., neutron $\beta$-decay $n\to p+e^{-}+\bar{\nu}_{e}$, and (iii) purely hadronic processes, e.g., $\Lambda\to p+\pi^{-}$. Perturbation theory is valid for weak and electromagnetic interactions. In the 1960s, a theory of electroweak interactions was developed by Sheldon Glashow, Abdus Salam and Steven Weinberg that can unify the electromagnetic and weak interactions.

So far, the experimental data on a wide range of leptonic and semi-leptonic processes are consistent with the assumption that the lepton fields enter the interaction only in the combinations
\begin{eqnarray}
J_{\alpha}(x)&=&\sum_{l}\bar{\psi}_{l}(x)\gamma_{\alpha}(1-\gamma_{5})\psi_{\nu_{l}}(x),\nonumber \\
J_{\alpha}^{\dagger}(x)&=&\sum_{l}\bar{\psi}_{\nu_{l}}(x)\gamma_{\alpha}(1-\gamma_{5})\psi_{l}(x),
\label{lc1}
\end{eqnarray}
where $J_{\alpha}(x)$ and $J_{\alpha}^{\dagger}(x)$ are called leptonic currents, $l=e, \mu, \tau$, $\psi_{l}$ and $\psi_{\nu_{l}}$ are  the corresponding quantized fields in Eq. (\ref{lc1}).  We can describe the weak interaction as due to the transmission of quanta, i.e., $W^{\pm}$. For example, the interaction Hamiltonian density of  quantum electrodynamics (QED), according to the intermediate vector boson (IVB) theory can be given by
\begin{equation}
{\cal H}_{I}(x)=g_{W}J^{\alpha\dagger}(x)W_{\alpha}(x)+g_{W}J^{\alpha}(x)W_{\alpha}^{\dagger}(x),
\label{lc2}
\end{equation}
where $g_{W}$ is a dimensionless coupling constant and the field $W_{\alpha}(x)$ describes the $W$ bosons in Eq. (\ref{lc2}). This interaction is known as a \lq V-A\rq interaction, since the current $J^{\alpha}(x)$ can be written as the difference of a vector part ($\gamma^{\mu}$) and an axial vector part ($\gamma^{\mu}\gamma^5$).

%%%%%%%%%%%%%%%%%%%%%%%%%%%%%%%%%%%%%%%%%%%%%%%%%%%%%%%%%%%%%%%%%%%%%%
%QCD-------------------------------------------------------
\section{QCD}

\hspace*{\parindent}Quantum chromodynamics (QCD) is the standard theory to describe the strong interactions, in which the color quantum number has been introduced as an extra degree of freedom. The color charge of a quark has three possible values, $r$, $g$ and $b$, while anti-quarks carry anti-colors, $\bar{r}$, $\bar{g}$ and $\bar{b}$. The mediating bosons of the quark-quark interactions are called \textit{gluons}, each carrying a color and an anti-color and postulated to belong to an octet of states.

Quarks and gluons are observed indirectly, which means that the evidence of their existence inside hadrons exists but these particles have not been observed singly. Experiments to study the strong interactions are performed with hadrons, not with the quarks and gluons that are described by quantum field theory (QFT). To explore or determine the quark and gluon structure of hadrons, structure functions are introduced to give the properties of a certain particle interaction without including all of the underlying physics. The experimental technique is to measure the angular distribution of some processes and compare it to that from a point particle, then the structure of the hadron can be deduced from some form factors(functions of the transferred momentum square). As an example, a charge distribution with electrons can be probed by measuring the cross section for scattering electrons:
\begin{equation}
\frac{d\sigma}{d\Omega}=(\frac{d\sigma}{d\Omega})_{point}|F(q)|^{2},
\end{equation}
where $q$ is the transferred momentum and $F(q)$ is the corresponding form factor.

%%%%%%%%%%%%%%%%%%%%%%%%%%%%%%%%%%%%%%%%%%%%%%%%%%%%%%%%%%%%%%%%%%%%%%%%
%formalism--------------------------------------------------------------
\doublespacing
\newpage
\chapter{FORMALISM}

\singlespacing
\hspace*{\parindent}The physics of the decay process $\lblc$ can be described by an effective Hamiltonian. In the presence of NP, the effective Hamiltonian for the quark-level transition $\bc$ can be written in the form \cite{eh01,eh02}
\begin{eqnarray}
{\cal{H}}_{eff}&=&\frac{G_{F}V_{cb}}{\sqrt{2}}\Big\{
\Big[\bar{c} \gamma_{\mu} (1-\gamma_{5})b  + g_{L}\bar{c}\gamma_{\mu} (1-\gamma_{5})b + g_{R}\bar{c} \gamma_{\mu}(1+\gamma_{5})b\Big]\bar{l}\gamma^{\mu}(1-\gamma_5)\nu_{l}\nonumber\\&& +  \Big[g_{S}\bar{c}b + g_{P}\bar{c} \gamma_{5} b\Big] \bar{l} (1-\gamma_5)\nu_l + h.c \Big\},
\label{eheqn}
\end{eqnarray}
where  $G_F = 1.1663787 \times 10^{-5} GeV^{-2}$ is the Fermi coupling constant, $V_{cb}$ is the Cabibbo-Kobayashi-Maskawa (CKM) matrix element, $g_{L,R,S,P}$ are NP couplings and I use $\sigma_{\mu \nu} = i[\gamma_\mu, \gamma_\nu]/2$. In this thesis, I have assumed the neutrinos to be always left chiral and the NP effect is mainly for the $\tau$ lepton. Here, I do not consider tensor operators in my work. Moreover, I do not assume any relation between $b \to u l^{-}\nu_l$ and $\bc$ transitions and hence the analysis does not include constraints from $B \to \tau \nu_{\tau}$. As it is expected, the SM  effective Hamiltonian corresponds to $g_L = g_R = g_S = g_P = 0$.  

In Refs.~\cite{bdnew3, bdnew4}, the authors had parametrized the NP in terms of the couplings $g_S$, $g_P$, $g_V=g_R+g_L$ and $g_A=g_R-g_L$ while in this thesis I have used $g_L$ and $g_R$ instead of $g_V$ and $g_A$ to align the analysis closer to realistic models \cite{bdls}. The couplings $g_{L,R,P}$ contribute to $R(D^*)$ while $g_{L,R,S}$ contribute to $R(D)$. The NP couplings are considered one at a time and the constraints on these couplings are obtained from $R(D^{(*)})$. 

\section{Decay Process}

\begin{figure}
\begin{center}
\includegraphics[width=15cm, height=6cm]{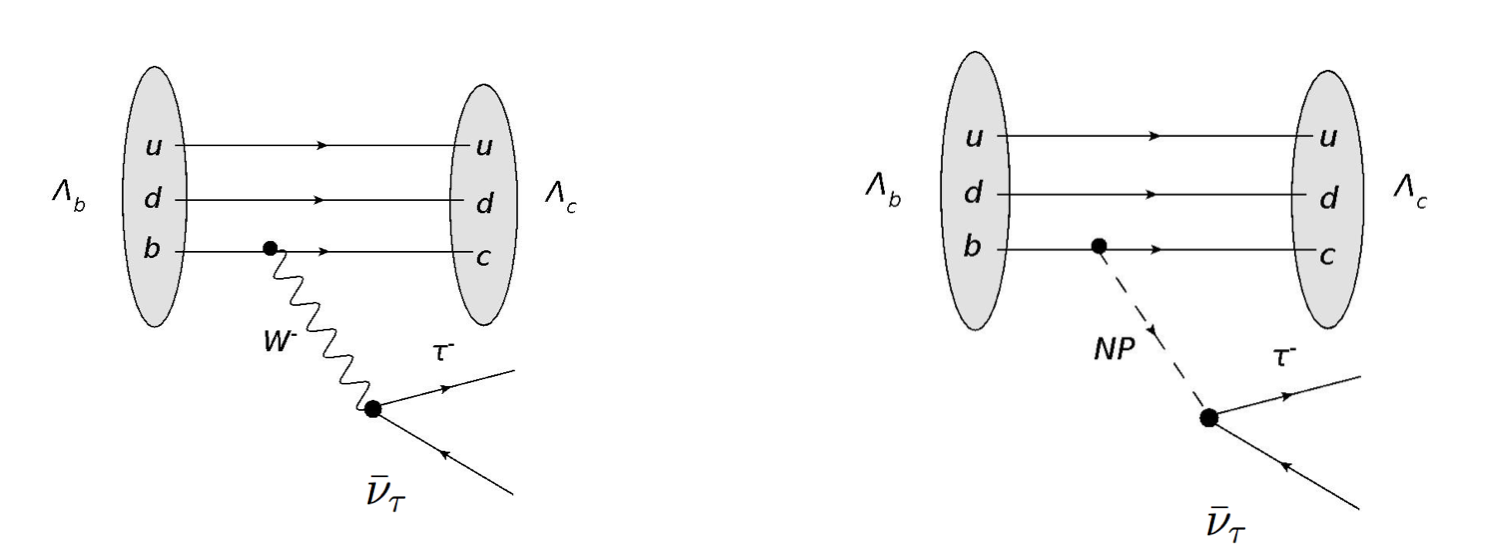}
\end{center}
\caption{$\Lambda_{b}$ Decay Process in the SM and with NP(some new mediating particles)}
\label{Fbdecay}
\end{figure}
\hspace*{\parindent}$\Lambda_{b}$ is a baryon of three quarks: $u$, $d$ and $b$, while $\Lambda_{c}$ is a baryon with $u$, $d$ and $c$ quarks. The semi-leptonic decay process under consideration is 
$$\Lambda_{b}(p)\to \tau^{-}(p_{1})+\bar{\nu_{\tau}}(p_{2})+\Lambda_{c}(p_{3}),$$
where $p$, $p_1$, $p_2$ and $p_3$ are four energy-momentum vectors respectively. Technically, the decay process $\lblc$ transits the \textit{b}-quark to the \textit{c}-quark, as is shown in Fig.~\ref{Fbdecay}. In the SM, the mediating boson is $W^{-}$, which will subsequently decay into a $\tau$ lepton and $\tau$ anti-neutrino. With NP, instead of $W^{-}$, the mediator can be a new particle, i.e., another new vector particle $W'^{-}$ or a scalar (or Higgs) particle $H^{-}$. I will consider these NP effects and compare the results with those from the SM.

\section{Partial Decay Rate}

\hspace*{\parindent}The partial decay rate of a particle of mass $m$ into $n$ bodies in its rest frame is given by 
\begin{eqnarray}
d\Gamma=\frac{(2\pi)^{4}}{2m}\lvert{M}\rvert^{2}d\Phi_{n}(p;p_{1},...,p_{n}),
\label{Edr}
\end{eqnarray}
where $M$ is the Feynman amplitude and $d\Phi_{n}$ is an element of $n$-body phase space given by
\bea
d\Phi_{n}(p;p_{1},...,p_{n})=\delta^{4}(p-\sum_{i=1}^{n}{p_{i}})\prod_{i=1}^{n}{\frac{d^{3}p_{i}}{(2\pi)^{3}2E_{i}}}.
\eea
This phase space element can be generated recursively
\bea
d\Phi_{n}(p;p_{1},...,p_{n})=d\Phi_{j}(q;p_{1},...,p_{j})\times d\Phi_{n-j+1}(p;q,p_{j+1},...,p_{n})(2\pi)^{3}dq^{2},
\eea
where $q^{2}=(\sum_{i=1}^{j}E_{i})^2-\lvert{\sum_{i=1}^{j}}p_{i}\rvert^{2}$. This form is particularly useful in the case where a particle decays into another particle that subsequently decays, e.g., $\Lambda_b \to \Lambda_c W^- \to \Lambda_c \tau^- \bar{\nu}_{\tau}$. A useful method to achieve the integration of $d\Gamma$ is given in Appendix \ref{Akinematics}.

\section{Feynman Amplitude}

\hspace*{\parindent}The Feynman amplitude $M$ includes all the physical processes. In fact, the key point to calculate the decay rate is to evaluate the $|M|^2$ appearing in Eq.~(\ref{Edr}).

The total Feynman amplitude here is
\begin{equation}
M_{total}=M_{SM}+M_{g_{L}}+M_{g_{R}}+M_{g_{S}}+M_{g_{P}},
\label{Efm}
\end{equation}
where $M_{g_{L,R,S,P}}$ are corresponding to the NP couplings $g_{L,R,S,P}$.

In the SM, the Feynman amplitude for this process is given by
\begin{equation}
M_{SM}=\frac{G_{F}V_{cb}}{\sqrt{2}}L^{\mu}H_{\mu},
\label{Efmsm}
\end{equation}
where the leptonic and hadronic currents are
\begin{eqnarray}
L^{\mu}&=&\bar{u}_{\tau}(p_{1})\gamma^{\mu}(1-\gamma_{5})v_{\nu_{\tau}}(p_{2}), \nonumber\\
H_{\mu} & = &\langle{\Lambda_c}|\bar{c}\gamma_{\mu}(1-\gamma_5)|{\Lambda_b}\rangle. \
\label{Elhcurrent}
\end{eqnarray}
The hadronic current is expressed in terms of six form factors,
\begin{eqnarray}
\langle{\Lambda_c}|\bar{c}\gamma_{\mu}b|{\Lambda_b}\rangle & = & \bar{u}_{\Lambda_{c}}(f_{1}\gamma_{\mu}+if_{2}\sigma_{\mu\nu}q^{\nu}+f_{3}q_{\mu})u_{\Lambda_{b}}, \nonumber\\
\langle{\Lambda_c}|\bar{c}\gamma_{\mu}\gamma_{5}b|\rangle{\Lambda_b} &= & \bar{u}_{\Lambda_{c}}(g_{1}\gamma_{\mu}\gamma_{5}+i g_{2}\sigma_{\mu\nu}q^{\nu}\gamma_{5}+g_{3}q_{\mu}\gamma_{5})u_{\Lambda_{b}}.
\label{hadronc}
\end{eqnarray}
Here $q= p-p_3$ is the transferred momentum and the form factors are functions of $q^2$.

When NP operators appear, we can obtain the hadronic current by considering the following relations:
\bea
q^{\mu}\langle\Lambda_{c}|\bar{c}\gamma_{\mu}b|\Lambda_{b}\rangle&=&q^{\mu}\bar{u}_{\lambda_{c}}(f_{1}\gamma_{\mu}+i f_{2}\sigma_{\mu\nu}q^{\nu}+f_{3}q_{\mu})u_{\lambda_{b}},\nn \\
q^{\mu}\langle\Lambda_{c}|\bar{c}\gamma_{\mu}\gamma_{5}b|\Lambda_{b}\rangle&=&q^{\mu}\bar{u}_{\lambda_{c}}(g_{1}\gamma_{\mu}\gamma_{5}+i g_{2}\sigma_{\mu\nu}q^{\nu}\gamma_{5}+g_{3}q_{\mu}\gamma_{5})u_{\lambda_{b}}.
\eea
Using the equations of motion, we'll finally get (the details are shown in Appendix \ref{npo})
\begin{eqnarray}
\langle{\Lambda_c}|\bar{c}b|{\Lambda_b}\rangle &= & \bar{u}_{\Lambda_{c}}(f_{1}\frac{\slashed{q}}{m_b-m_{c}}+f_{3}\frac{q^2}{m_b-m_{c}})u_{\Lambda_{b}}, \nonumber\\
\langle{\Lambda_c}|\bar{c}\gamma_{5}b|{\Lambda_b}\rangle &= &\bar{u}_{\Lambda_{c}}(-g_{1}\frac{\slashed{q}\gamma_{5}}{m_b+m_{c}}-g_{3}\frac{q^2\gamma_{5}}{m_b+m_{c}})u_{\Lambda_{b}}.
\end{eqnarray}
where $m_b$ and $m_c$ are the masses of the $b$ quark and $c$ quark.

To obtain the corresponding unpolarized decay rate, we need to average $|M|^2$ over all initial polarization states and sum it over all final polarization states,
\bea
\bar{\lvert{M}\rvert^{2}}&=&\frac{1}{2}\sum_{spin}{\lvert{M}\rvert^{2}}\nn\\&=&
\frac{G_{f}^{2}\lvert{V_{cb}}\rvert^{2}}{4}L^{\mu\nu}H_{\mu\nu},
\label{Em2}
\eea
where $L^{\mu\nu}$ stands for the leptonic part and $H_{\mu\nu}$ stands for the hadronic part ($L^{\mu\nu}$ and $H_{\mu\nu}$ are tensors in the SM and with a vector NP effect). In the following part of this section, I will give the details of how to obtain $\bar{\lvert{M}\rvert^{2}}$ both in the SM and with NP effects. However, from Eq. (\ref{Efm}) we know that there should be some crossing terms between the SM and NP effects. We are not going to consider the crossing term between two different NP effects since we just consider one NP coupling at a time. To get the final result form of Eq.~(\ref{Em2}), we have to consider the kinematics of the decay process. The kinematics here is considered in the rest frame of $\Lambda_b$, and details are given in Appendix \ref{Akinematics}.

\subsection{SM}

\hspace*{\parindent}In the SM, the leptonic tensor in Eq.~(\ref{Em2}) is
\begin{eqnarray}
\sum_{spin}L^{\mu\nu}_{SM}&=&\sum_{spin}{[\bar{u}_{\tau}(p_{1})\gamma^{\mu}(1-\gamma_{5})v_{\nu_{\tau}}(p_{2})][\bar{v}_{\nu_{\tau}}(p_{2})\gamma^{\nu}(1-\gamma_{5})u_{\tau}(p_{1})]}\nn\\&=&Tr[(\slashed{p_{1}}+m_{1})\gamma^{\mu}(1-\gamma_{5})(\slashed{p_{2}}-m_2)\gamma^{\nu}(1-\gamma_{5})]\nn\\&=&8(-g^{\mu\nu}p_{1}\cdot p_{2}-i\epsilon^{\mu\nu\rho\sigma}p_{1\rho}p_{2\sigma}+p_{1}^{\mu}p_{2}^{\nu}+p_{1}^{\nu}p_{2}^{\mu}).
\label{Eltsm}
\end{eqnarray}
Here, $m_1$ is the mass of $\tau$ lepton and $m_2$ is the mass of tau neutrino. I already treat the neutrino as massless, $m_2\rightarrow 0$.
The hadronic tensor in Eq.~(\ref{Em2}) in the SM is
\begin{eqnarray}
\sum_{spin}{H_{SM}}_{\mu\nu}&=&\sum_{spin}{H_{SM}}_{\mu}{H_{SM}}_{\nu}^{*}\nn\\&=&\sum_{spin}(A-B)(A^{*}-B^{*})\nn\\&=&\sum_{spin}(AA^{*}-AB^{*}-BA^{*}+BB^{*}),
\label{Ehsm}
\end{eqnarray}
where
\bea
A&=&\bar{u}_{\lambda_{c}}(f_{1}\gamma_{\mu}+i f_{2}\sigma_{\mu\nu}q^{\nu}+f_{3}q_{\mu})u_{\lambda_{b}},\nn\\
B&=&\bar{u}_{\lambda_{c}}(g_{1}\gamma_{\mu}\gamma_{5}+i g_{2}\sigma_{\mu\nu}q^{\nu}\gamma_{5}+g_{3}q_{\mu}\gamma_{5})u_{\lambda_{b}},\nn\\
A^{*}&=&\bar{u}_{\lambda_{b}}(f_{1}\gamma_{\nu}-i f_{2}\sigma_{\nu\delta}q^{\delta}+f_{3}q_{\nu})u_{\lambda_{c}},\nn\\
B^{*}&=&\bar{u}_{\lambda_{b}}(g_{1}\gamma_{\nu}\gamma_{5}+i g_{2}\sigma_{\nu\delta}q^{\delta}\gamma_{5}-g_{3}q_{\nu}\gamma_{5})u_{\lambda_{c}}.
\label{Eabab}
\eea

\subsection{NP Effects}

\hspace*{\parindent}Let's consider the NP effects for one NP coupling at a time and set the others to zero. For the vector NP effects, we consider the case with only $g_L$ present and the case with only $g_R$ present. For the scalar/pseudoscalar NP effects, we consider the case with only $g_S$ or $g_P$ present.

For the vector NP effect with only $g_L$ present, the NP coupling will appear in the hadronic current part in the Feynman amplitude and the leptonic current part is the same as that in the SM. Therefore, the leptonic tensor $L^{\mu\nu}_{g_{L}}=L^{\mu\nu}_{SM}$ has the form in Eq.~(\ref{Eltsm}). The hadronic tensor with only $g_L$ present is
\bea
{H_{SM+g_L}}_{\mu\nu}&=&\sum_{spin}{H_{SM+g_L}}_{\mu}{H_{SM+g_L}}^{*}_{\nu}\nn\\&=&\sum_{spin}({H_{SM}}_{\mu}{H_{SM}}_{\nu}^{*}+{H_{SM}}_{\mu}{H_{g_L}}_{\nu}^{*}+{H_{g_L}}_{\mu}{H_{SM}}_{\nu}^{*}+{H_{g_L}}_{\mu}{H_{g_L}}_{\nu}^{*})\nn\\&=&\sum_{spin}(1+g_L^*+g_L+|g_L|^2)(A-B)(A^{*}-B^{*}).
\label{EgL}
\eea
With the $g_{R}$ present, we will have $L^{\mu\nu}_{g_{R}}=L^{\mu\nu}_{SM}$  for the same reason as with $g_L$. The corresponding hadronic tensor is
\bea
{H_{SM+g_R}}_{\mu\nu}&=&\sum_{spin}{H_{SM+g_R}}_{\mu}{H_{SM+g_R}}^{*}_{\nu}\nn\\&=&\sum_{spin}({H_{SM}}_{\mu}{H_{SM}}_{\nu}^{*}+{H_{SM}}_{\mu}{H_{g_R}}_{\nu}^{*}+{H_{g_R}}_{\mu}{H_{SM}}_{\nu}^{*}+{H_{g_R}}_{\mu}{H_{g_R}}_{\nu}^{*})\nn\\&=&\sum_{spin}[(A-B)(A^{*}-B^{*})+g_R^{*}(A-B)(A^{*}+B^{*})\nn\\&&+g_R (A+B)(A^{*}-B^{*})+|g_R|^2 (A+B)(A^{*}+B^{*})].
\label{EgR}
\eea
The $A$, $B$, $A^*$ and $B^*$ appearing in Eq.~(\ref{EgL}) and Eq.~(\ref{EgR}) are given in Eq.~(\ref{Eabab}).

For the scalar and pseudoscalar NP effects, the Feynman amplitudes are given by
\bea
M_{g_{S}}&=&g_{S}\frac{G_{F}V_{cb}}{\sqrt{2}}L_{S}C,\nn\\
M_{g_{P}}&=&g_{P}\frac{G_{F}V_{cb}}{\sqrt{2}}L_{S}D,
\label{EgSP}
\eea
where $L_{S}$ is the leptonic current, $C$ and $D$ are the corresponding hadronic currents. Then the $\bar{|M|}^2$ in Eq.~(\ref{Em2}) 
with NP effect with only $g_S$ present becomes
\bea
{\bar{\lvert{M}\rvert^{2}}}_{SM+g_S}&=&\frac{1}{2}\sum_{spin}({\lvert{M_{SM}}\rvert^{2}}+M_{SM}M_{g_S}^{*}+M_{g_S}M_{SM}^{*}+|M_{g_S}|^2)\nn\\&=&\frac{G_{f}^{2}\lvert{V_{cb}}\rvert^{2}}{4}\sum_{spin}[LL^{*}(A-B)(A^{*}-B^{*})+g_{S}^{*}LL_{S}^{*}(A-B)C^{*}\nn\\&&+g_{S}L_{S}L^{*}C(A^*-B^*)+|g_{S}|^{2}L_{S}L_{S}^{*}CC^*],
\eea
and the  $\bar{|M|}^2$ with only $g_P$ present becomes
\bea
{\bar{\lvert{M}\rvert^{2}}}_{SM+g_P}&=&\frac{1}{2}\sum_{spin}({\lvert{M_{SM}}\rvert^{2}}+M_{SM}M_{g_P}^{*}+M_{g_P}M_{SM}^{*}+|M_{g_P}|^2)\nn\\&=&\frac{G_{f}^{2}\lvert{V_{cb}}\rvert^{2}}{4}\sum_{spin}[LL^{*}(A-B)(A^{*}-B^{*})+g_{P}^{*}LL_{S}^{*}(A-B)D^{*}\nn\\&&+g_{P}L_{S}L^{*}D(A^*-B^*)+|g_{P}|^{2}L_{S}L_{S}^{*}DD^*].
\eea
Here, $L$ is the leptonic current in Eq.~(\ref{Elhcurrent}) and $L^*$ is its conjugate part. In addition, 
\bea
L_S&=&\bar{u}_{\tau}(p_{1})(1-\gamma_{5})v_{\nu_{\tau}}(p_{2}),\nn\\
C&=&\bar{u}_{\lambda_{c}}(f_{1}\frac{\slashed{q}}{m_{b}-m_{c}}+f_{3}\frac{q^{2}}{m_{b}-m_{c}})u_{\lambda_{b}},\nn\\
D&=&\bar{u}_{\lambda_{c}}(-g_{1}\frac{\slashed{q}\gamma_{5}}{m_{b}+m_{c}}-g_{3}\frac{q^{2}\gamma_{5}}{m_{b}+m_{c}})u_{\lambda_{b}},
\eea
and their corresponding conjugate parts are 
\bea
L_{S}^{*}&=&\bar{v}_{\nu_{\tau}}(p_{2})(1+\gamma_{5})u_{\tau}(p_{1}),\nn\\
C^{*}&=&\bar{u}_{\lambda_{b}}(f_{1}\frac{\slashed{q}}{m_{b}-m_{c}}+f_{3}\frac{q^{2}}{m_{b}-m_{c}})u_{\lambda_{c}},\nn\\
D^{*}&=&\bar{u}_{\lambda_{b}}(-g_{1}\frac{\slashed{q}\gamma_{5}}{m_{b}+m_{c}}+g_{3}\frac{q^{2}\gamma_{5}}{m_{b}+m_{c}})u_{\lambda_{c}}.
\eea

The cross terms between two different NP couplings are zero since we consider one NP coupling at a time, as Eq.~(\ref{Efm}) indicates.

\section{Observables}
\label{Sob}

\hspace*{\parindent}The calculation is based on integration of Eq.~(\ref{Edr}), which gives us the decay rate of the process $\lblc$ directly. In this thesis, we'll define the following observables.
\bea
R_{\Lambda_{b}} & = & \frac{BR[\lblc]}{BR[\bcl]}.
\label{Eratio1}
\eea
Here $\ell$ represents $\mu$ or $e$. The branching ratio $BR$ for a specific decay process is defined by
\bea
BR_{i}&=&\frac{\Gamma_{i}}{\sum \Gamma_{i}},
\eea
where $\Gamma_{i}$ is the decay rate for this process and $\sum{\Gamma_{i}}$ is the total decay rate.

The differential distributions with respect to the transferred momentum square $q^2$ will be shown in the results ($d\Gamma/d q^2$). Also, we will define the ratio of differential distributions
\bea
B_{\Lambda_b}(q^2) & = & \frac{d\Gamma[\lblc]}{d q^2}\big/\frac{ d \Gamma[\bcl]}{d q^2}.
\label{Eratio2}
\eea

The results will show that these observables are not very sensitive to variations in the hadronic form factors.

%%%%%%%%%%%%%%%%%%%%%%%%%%%%%%%%%%%%%%%%%%%%%%%%%%%%%%%%%%%%%%%%%%%%%%%%%
%numerical results------------------------------------------------------
\doublespacing
\chapter{NUMERICAL RESULTS}

\singlespacing
\hspace*{\parindent}In this section, I will present the constraints on the NP couplings, then I will discuss and show the form factors used in this work. Finally, I will present the result graphs of the observables defined in section \ref{Sob}.

\section{NP Couplings}

\begin{figure}
\begin{center}
\includegraphics[width=8cm, height=5cm]{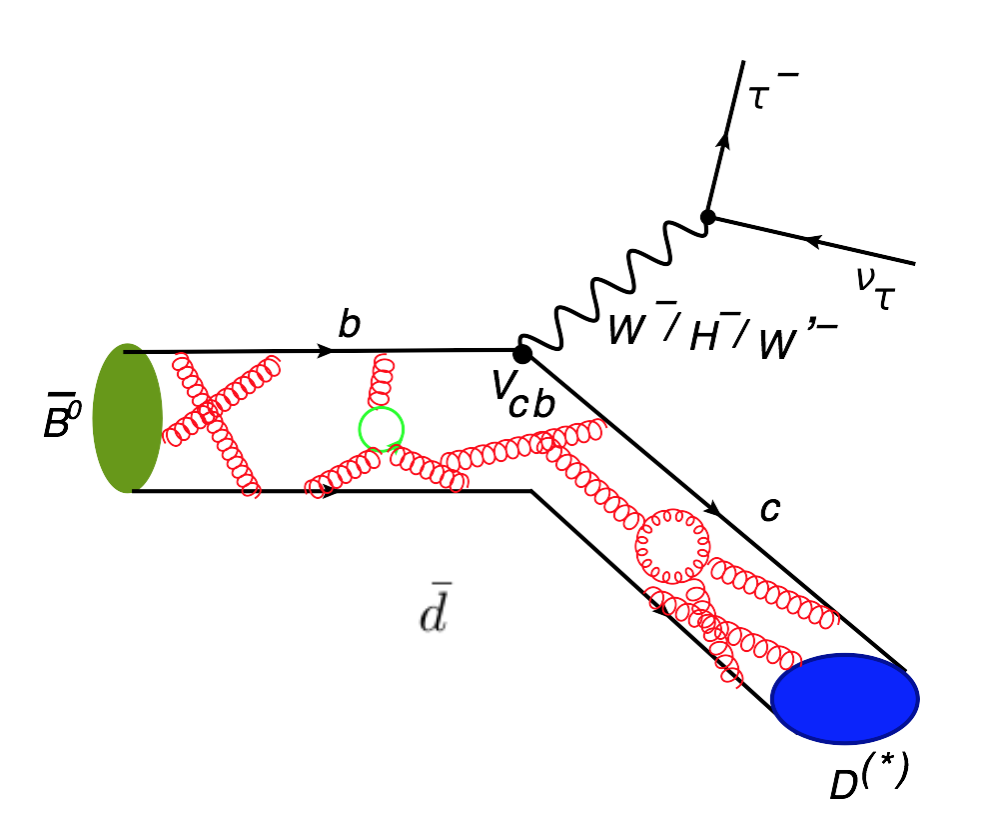}
\end{center}
\caption{$\bar{B}\to D^{(*)}\tau^{-}\bar{\nu}_{\tau}$ Decay Process}
\label{Fbddecay}
\end{figure}

\hspace*{\parindent}The NP constraints on the NP couplings are obtained from $R(D^{(*)})$ in Eq.~(\ref{babar}) and Eq.~(\ref{bdsm}). The relative decay processes are
\begin{center}
$\bar{B}\to D\tau^{-}\bar{\nu}_{\tau}$\ , and \  $\bar{B}\to D^*\tau^{-}\bar{\nu}_{\tau}$.
\end{center}
The two main reasons we use the NP constraints from the decay process $\bar{B}\to D^{(*)}\tau^{-}\bar{\nu}_{\tau}$ here are: the experimental results of $R(D^{(*)})$ from the BaBar Collaboration deviating from those in SM provide a hint of NP \cite{BaBar1,BaBar2}, and both $\Lambda_b$ baryon decay and $\bar{B}$ meson decay involve the transition $\bc$, as shown in Fig.~\ref{Fbdecay} and in Fig.~\ref{Fbddecay}. The formalism to constrain the NP couplings here is 
\begin{equation}
R_{exp}=R_{SM}\big (1+g_{NP}\frac{M_{NP}M_{SM}^{*}}{|M_{SM}|^2}+g_{NP}^{*}\frac{M_{SM}M_{NP}^{*}}{|M_{SM}|^2}+|g_{NP}|^{2}\frac{|M_{NP}|^2}{|M_{SM}|^2}\big ),
\label{Enp}
\end{equation}
where $g_{NP}$ stands for $g_{L,R,S,P}$.

The NP coupling $g_S$ only contributes to $R(D)$ and $g_P$ only contributes to $R(D^*)$ while  $g_{L,R}$  contributes to both $R(D)$  and $R(D^*)$. These couplings can be derived from Eq.(\ref{Enp}) and their allowed regions are shown in Fig. \ref{Fnp}. To get them, we need the form factors of $\bar{B}\to D^{(*)}$ decay, given in the Appendix \ref{Abdsdecay} and \ref{Abddecay}. 

\begin{figure}
\begin{center}
\includegraphics[width=6.5cm, height=5.5cm]{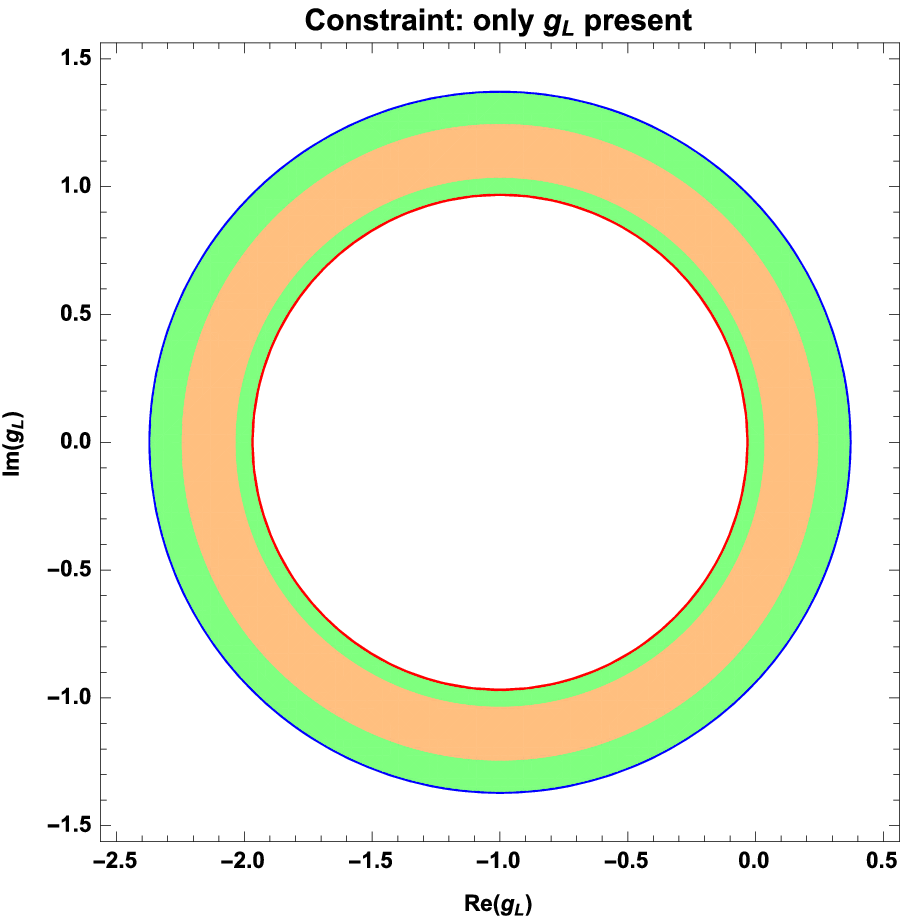}
\includegraphics[width=6.5cm, height=5.5cm]{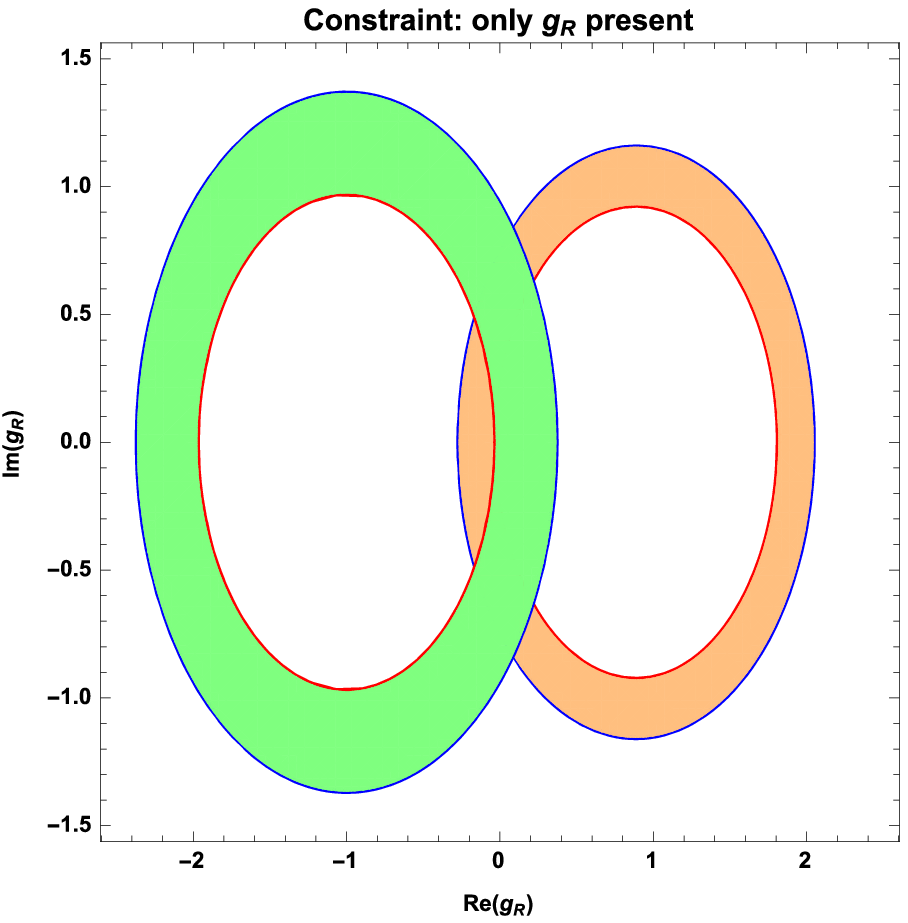}
\includegraphics[width=6.5cm, height=5.5cm]{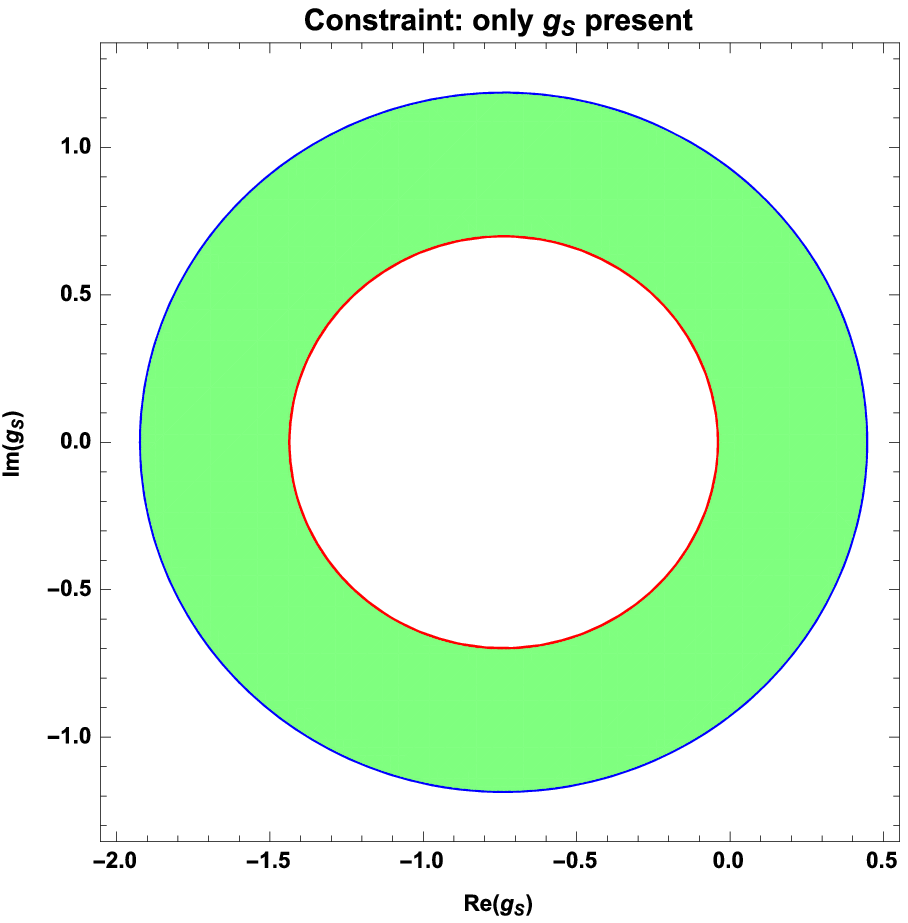}
\includegraphics[width=6.5cm, height=5.5cm]{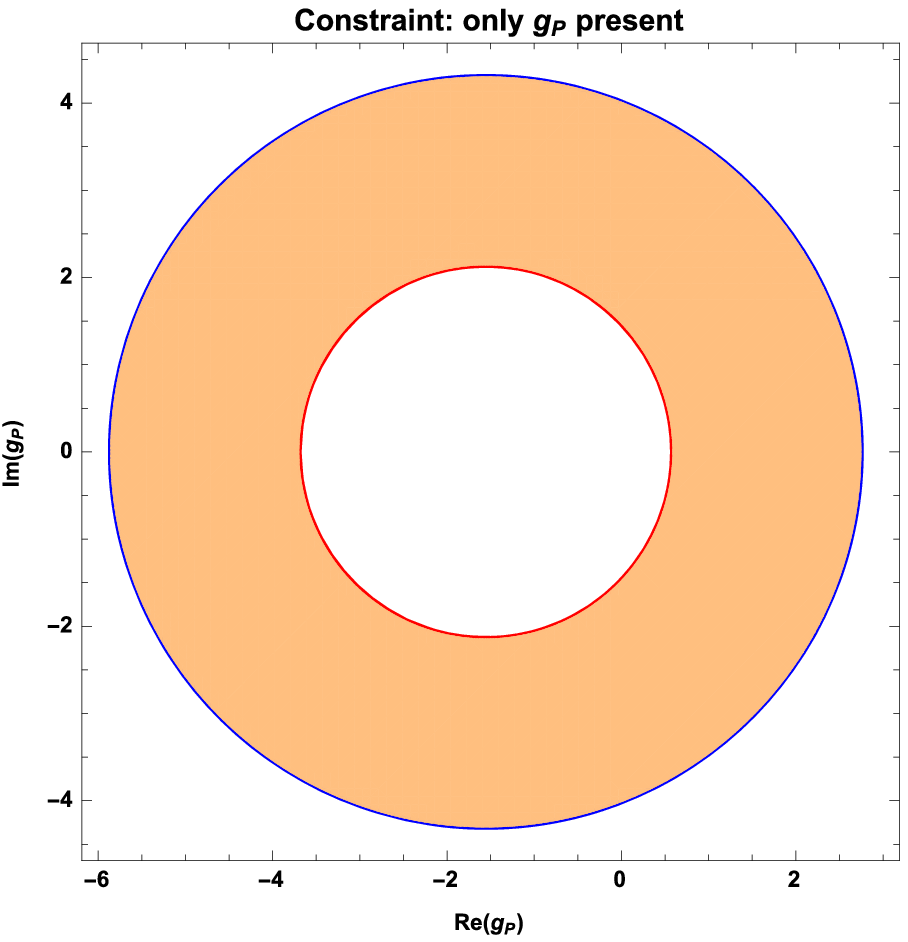}
\end{center}
\caption{The figures show the constraints on the NP couplings taken one at a time at the 95\% CL limit. When the 
couplings contribute to both $R(D)$  and $R(D^*)$ the green areas indicate constraints from $R(D)$ and the orange ones from
$R(D^*)$. }
\label{Fnp}
\end{figure}

\section{Form Factors}

\hspace*{\parindent}One of the main inputs in our calculations are the form factors. As first principle, lattice calculations of the form factors are not yet available. The form factors we use here are from QCD sum rules, which is a well known approach to compute non-perturbative effects like form factors for systems with both light and heavy quarks \cite{FF1,FF2}. Another approach to investigate the heavy quark systems involving quark flavour and spin symmetry ,called Heavy Quark Effective Theory (HQET) \cite{FF3}. The form factors we choose are consistent with HQET.

There are different values of the relative parameters for the kinematic region ascribed to the continuum model. This gives us some choices of the form factors. In Ref.~\cite{FF2} various parametrizations of the form factors are given. They are shown below.

\begin{table}[tbh]
\center
\begin{tabular}{|c|c|c|c|}\hline
$continuum \ model$ & $\kappa$ & $F_1^V(t)=f_1$ & $F_2^V(t)(GeV^{-1})=f_2$\\
\hline
$rectangular$ & 1 & ${6.66/(20.27-t)}$ & ${-0.21/(15.15-t)}$ \\
$rectangular$ & 2 & ${8.13/(22.50-t)}$ & ${-0.22/(13.63-t)}$\\
$triangular$ & 3 & ${13.74/(26.68-t)}$ & ${-0.41/(18.65-t)}$\\
$triangular$ & 4 & ${16.17/(29.12-t)}$ & ${-0.45/(19.04-t)}$\\
\hline
\end{tabular}
\caption{Various Choices of Form Factors ($t=q^2$)}
\label{FF}
\end{table}

The form factors in Table \ref{FF} are based on four continuum models indicated by $\kappa=1, 2, 3, 4$. However, the differences of the results from these models are very small, which can be shown in the result graphs. Moreover, these form factors in Table \ref{FF} satisfy the HQET relation in the $m_b \to \infty$ limit. They have the following relations:
\begin{equation}
f_{1}=g_{1}, \quad f_{2}=g_{2}, \quad f_{3}=g_{3}=0.\
\end{equation}

%%%%%%%%%%%%%%%%%%%%%%%%%%%%%%%%%%%%%%%%%%%%%%%%%%%%%%%%%%%%%%%%%%%%
\section{Result Graphs}

\hspace*{\parindent}I have used the following masses in my calculations. The masses of the particles are $m= 5.6195$ GeV, $m_{\tau}= 1.77682$ GeV,  $m_{\mu}=0.10565837 $ GeV, $m_{3} = 2.28646$ GeV, $m_b = 4.66$ GeV and $m_c = 1.275$ GeV \cite{pdg}. In the following I will present the results for observables $R_{\lb}$, $d\Gamma/dq^2$ and $B_{\lb}(q^2)$. For the first and third observables I use different models of the form factors given in Table \ref{FF}. For the differential distribution $d\Gamma/dq^2$, I present the average results over the form factors.

\begin{table}[tbh]
\center
\begin{tabular}{|c|c|c|c|c|c|c|c|}\hline
continuum \ model & 1 & 2 & 3 & 4 & Average & Ref.~\cite{Gutsche:2015mxa} & Ref.~\cite{Woloshyn}\\
\hline
$R_{\Lambda_{b}}(SM)$ & 0.31  & 0.29 & 0.28 & 0.27 & 0.29 &0.29 &0.31\\
\hline
\end{tabular}
\caption{Values of $R_{\Lambda_{b}}$ in the SM}
\label{SMpredictions}
\end{table}

In Table \ref{SMpredictions}, the prediction for $R_{\lb}$ in the SM are given for the various choices of the form factors in Table \ref{FF}.  I also compare our results with other calculations of this quantity
by other groups using different form factors. The average value we found for $R_{\lb}$ in the SM is  $R_{\lb,SM}= 0.29$. This agrees very well with values for this quantity obtained in Ref.~\cite{Gutsche:2015mxa}, which uses a covariant confined quark model for the form factors,
 and Ref.~\cite{Woloshyn} which uses the form factor model in Ref.~\cite{Pervin:2005ve}. These results indicate that the ratio $R_{\lb}$ is largely free from form factor uncertainties making it an excellent probe to find new physics.

\begin{figure}
\begin{center}
\includegraphics[width=6.5cm, height=5.5cm]{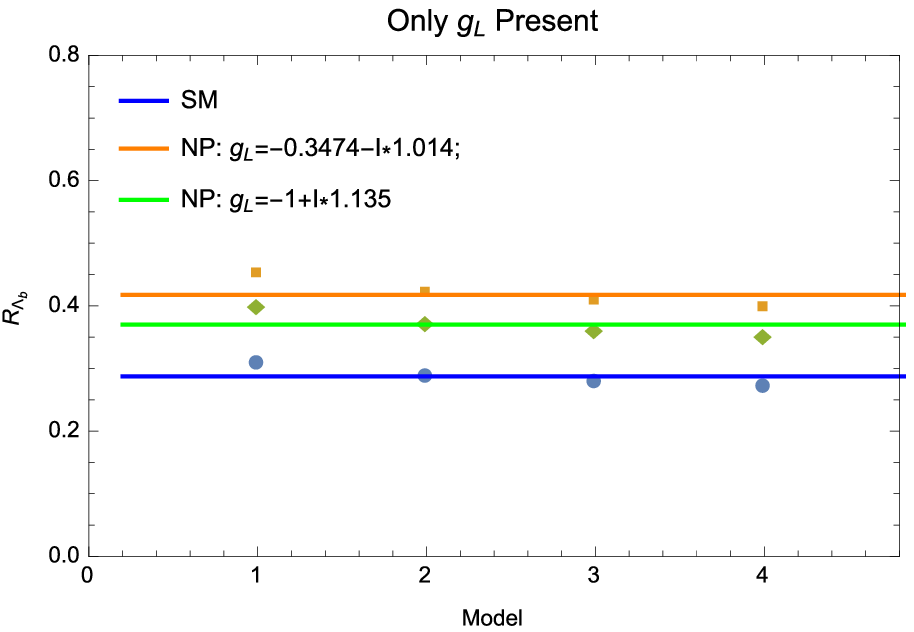}
\includegraphics[width=6.5cm, height=5.5cm]{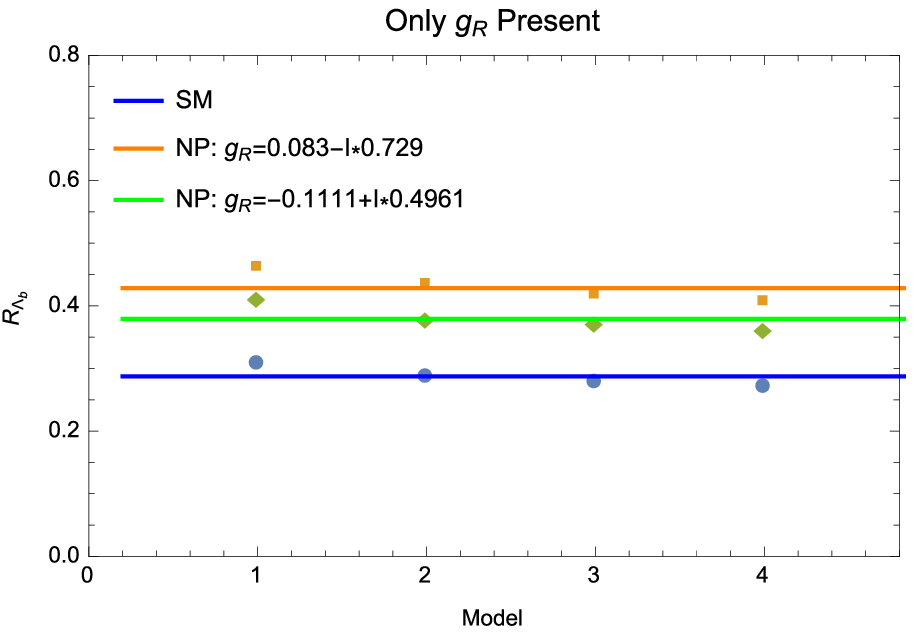}
\includegraphics[width=6.5cm, height=5.5cm]{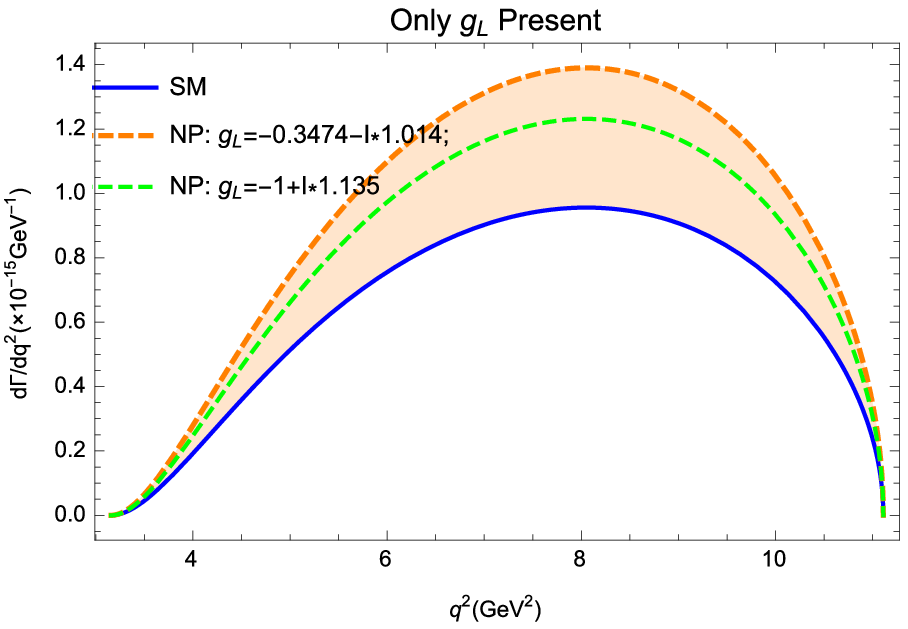}
\includegraphics[width=6.5cm, height=5.5cm]{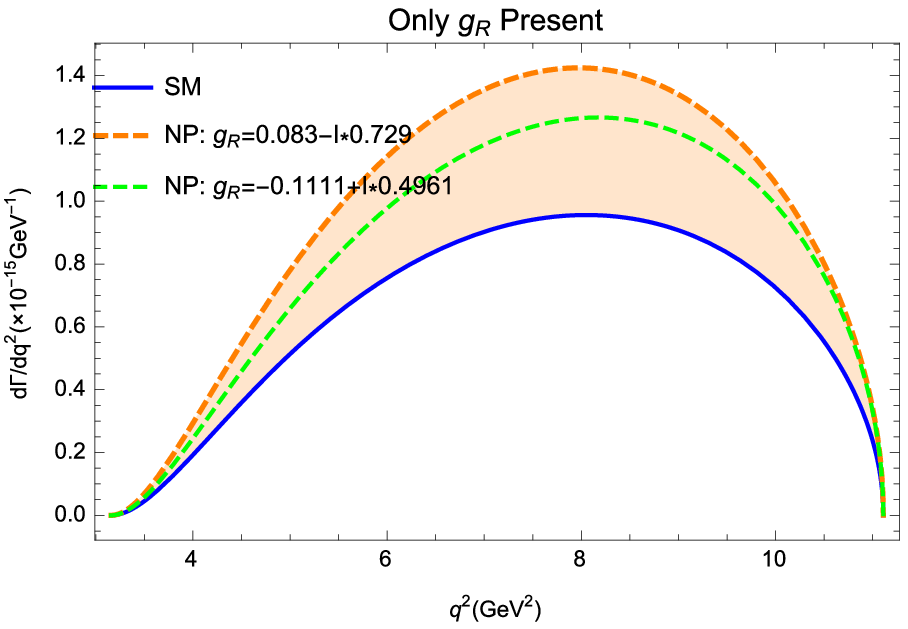}
\includegraphics[width=6.5cm, height=5.5cm]{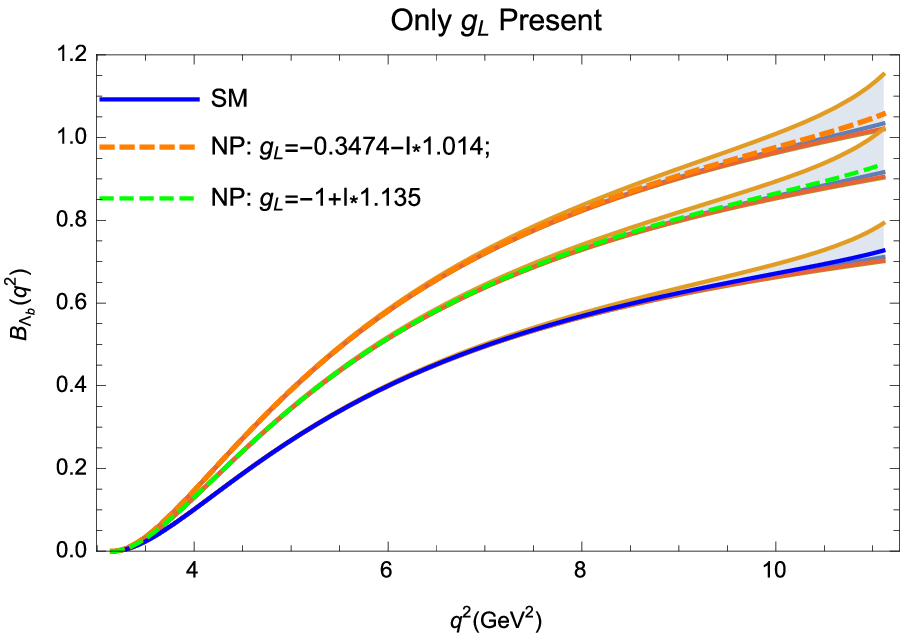}
\includegraphics[width=6.5cm, height=5.5cm]{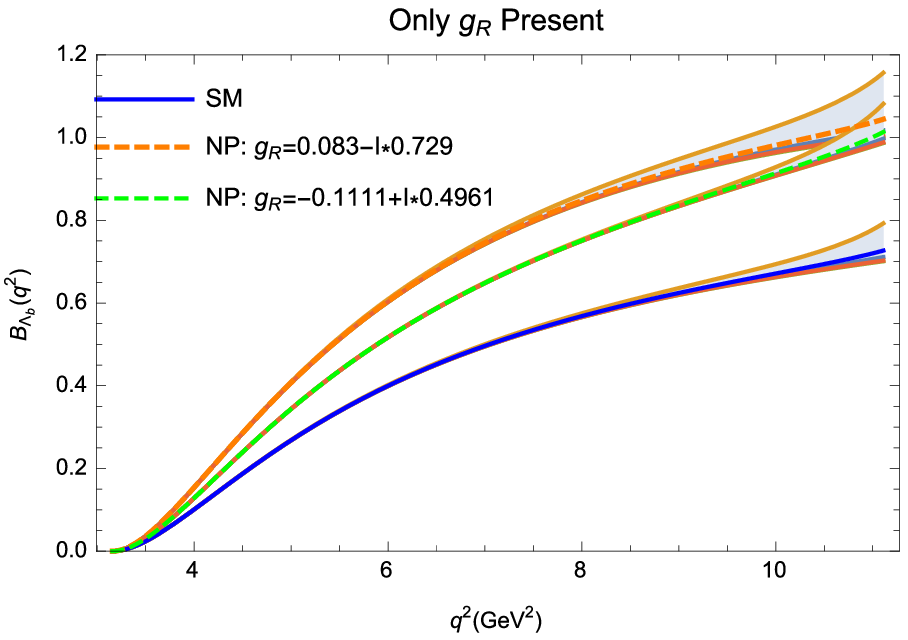}
\end{center}
\caption{The graphs on the left-side (right-side) show the compared results between the standard model and new physics with only $g_L$ ($g_R$) present.  The top and bottom row of graphs depict
$R_{\lb}=BR[{\Lambda_{b}\to\Lambda_{c}\tau\bar{\nu}_{\tau}}]/BR[{\Lambda_{b}\to\Lambda_{c}\ell\bar{\nu}_{\ell}}]$ 
and the ratio of differential distributions $B_{\lb}(q^2)=\frac{d\Gamma}{dq^{2}}(\Lambda_{b}\to\Lambda_{c}\tau\bar{\nu}_{\tau})/\frac{d\Gamma}{dq^{2}}(\Lambda_{b}\to\Lambda_{c}\ell\bar{\nu}_{\ell})$ as a function of $q^{2}$,
respectively for the various form factors in Table \ref{FF}. The middle graphs depict 
the average differential decay rate with respect to $q^2$ for  the process $\Lambda_{b}\to\Lambda_{c}\tau\bar{\nu}_{\tau}$. Some representative values of the couplings have been chosen. 
}
\label{gL}
\end{figure}

\begin{figure}
\begin{center}
\includegraphics[width=6.5cm, height=5.5cm]{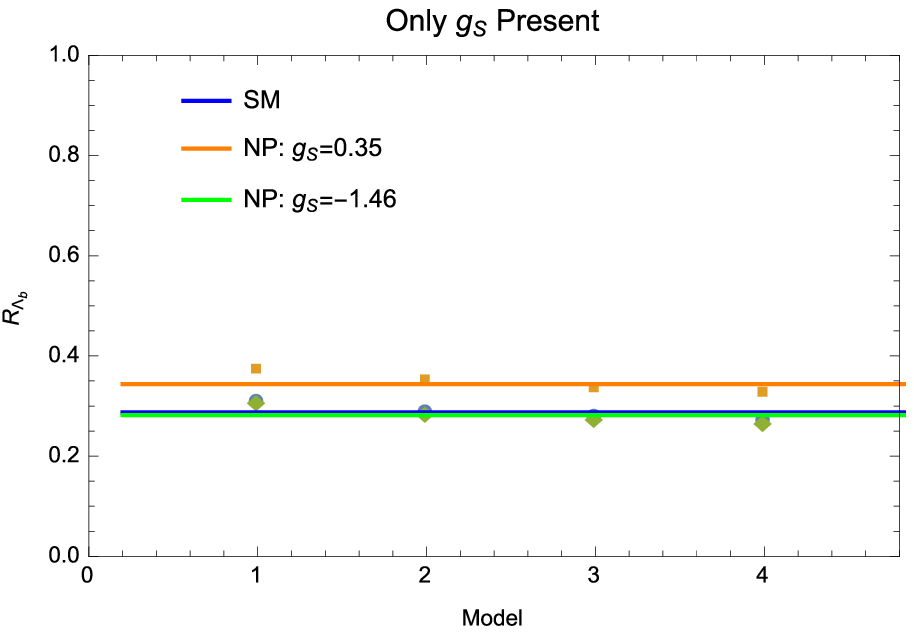}
\includegraphics[width=6.5cm, height=5.5cm]{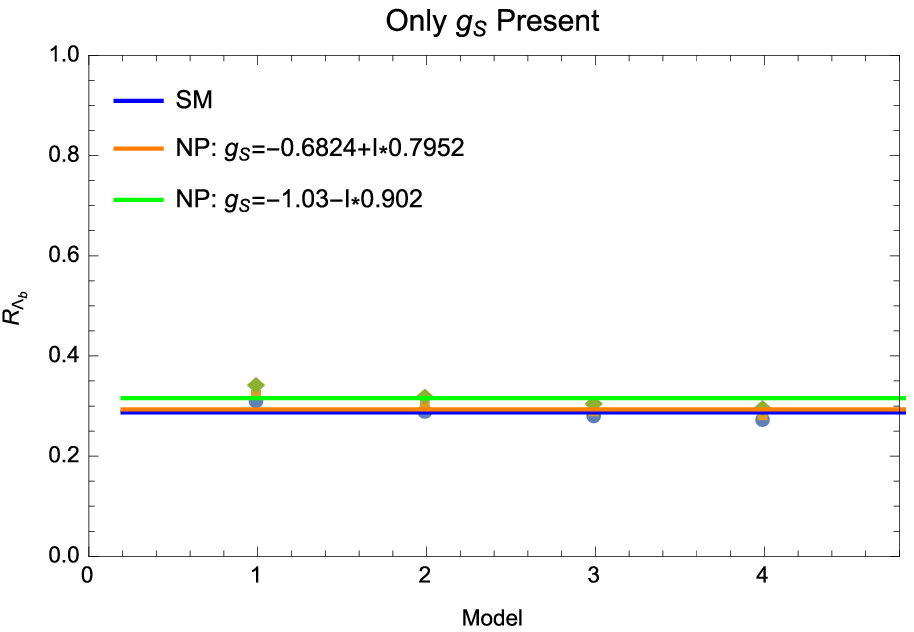}
\includegraphics[width=6.5cm, height=5.5cm]{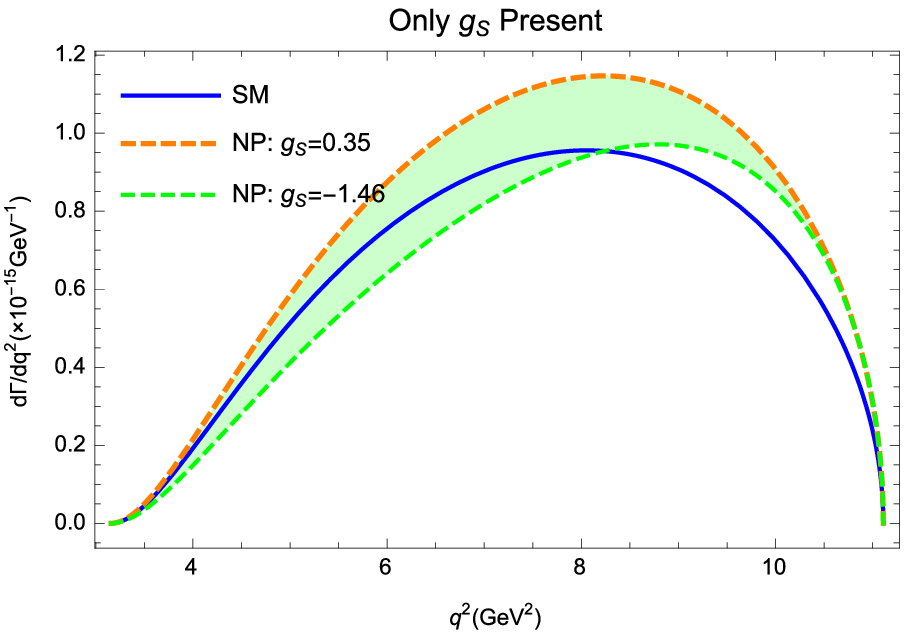}
\includegraphics[width=6.5cm, height=5.5cm]{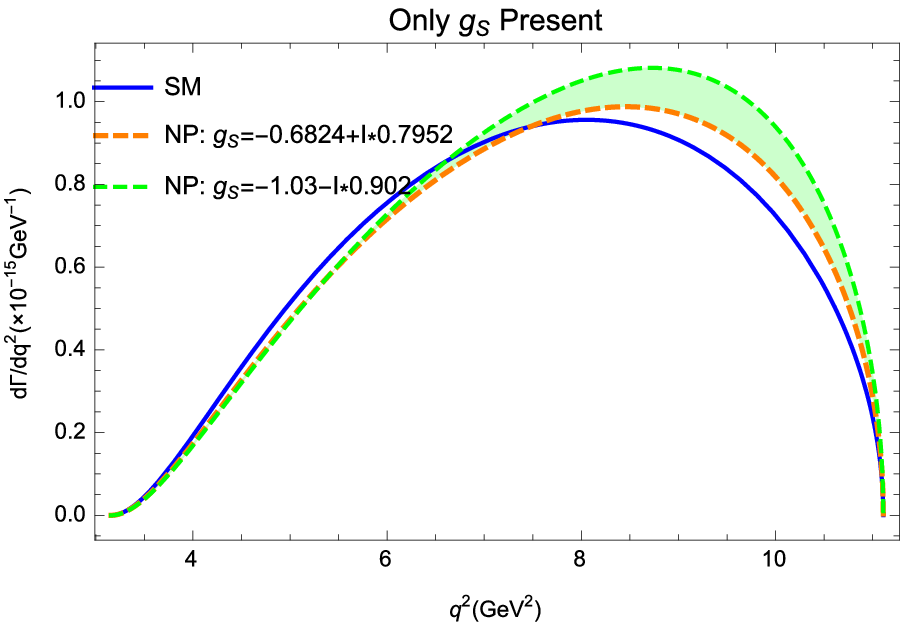}
\includegraphics[width=6.5cm, height=5.5cm]{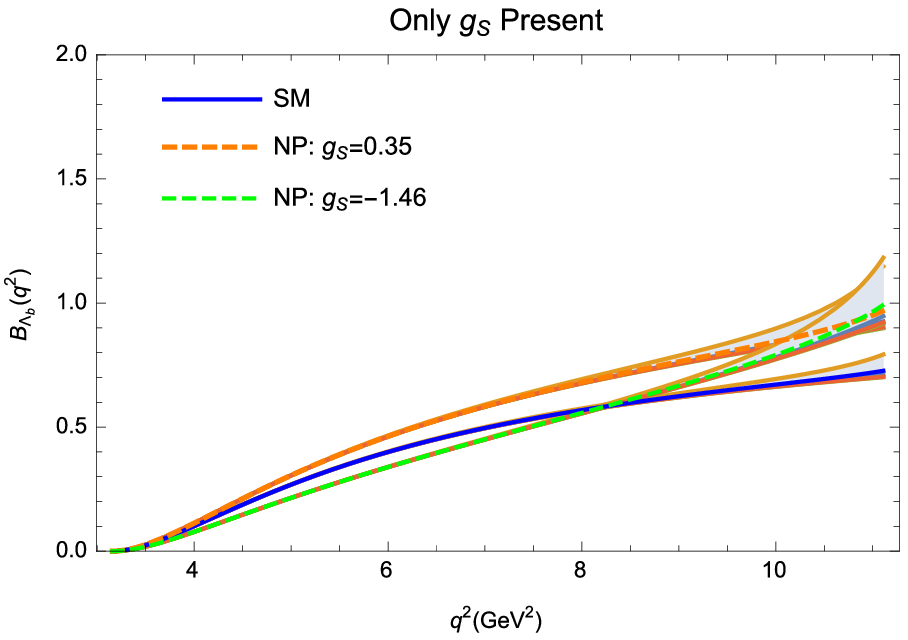}
\includegraphics[width=6.5cm, height=5.5cm]{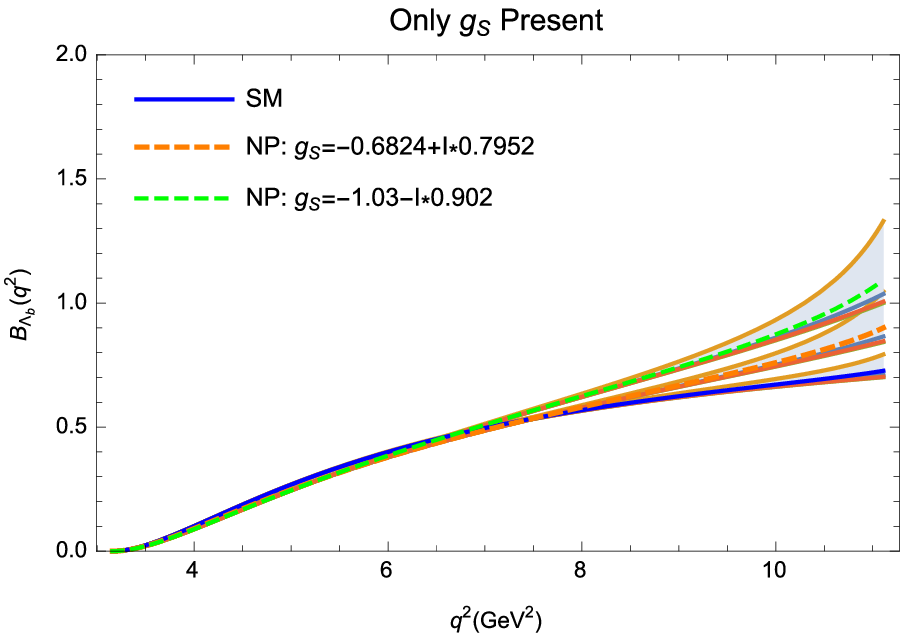}
\end{center}
\caption{
The figures show the compared results between the standard model and new physics with only $g_{S}$ present.
The top and bottom row of graphs depict
$R_{\lb}=BR[{\Lambda_{b}\to\Lambda_{c}\tau\bar{\nu}_{\tau}}]/BR[{\Lambda_{b}\to\Lambda_{c}\ell\bar{\nu}_{\ell}}]$ 
and the
ratio of differential distributions $ B_{\lb}(q^2)=\frac{d\Gamma}{dq^{2}}(\Lambda_{b}\to\Lambda_{c}\tau\bar{\nu}_{\tau})/\frac{d\Gamma}{dq^{2}}(\Lambda_{b}\to\Lambda_{c}\ell\bar{\nu}_{\ell})$ as a function of $q^{2}$,
respectively for the various form factors in Table \ref{FF}. The middle graphs depict 
the average differential decay rate with respect to $q^2$ for  the process $\Lambda_{b}\to\Lambda_{c}\tau\bar{\nu}_{\tau}$. Some representative values of the couplings have been chosen. 
}
\label{gS1}
\end{figure}

\begin{figure}
\begin{center}
\includegraphics[width=6.5cm, height=5.5cm]{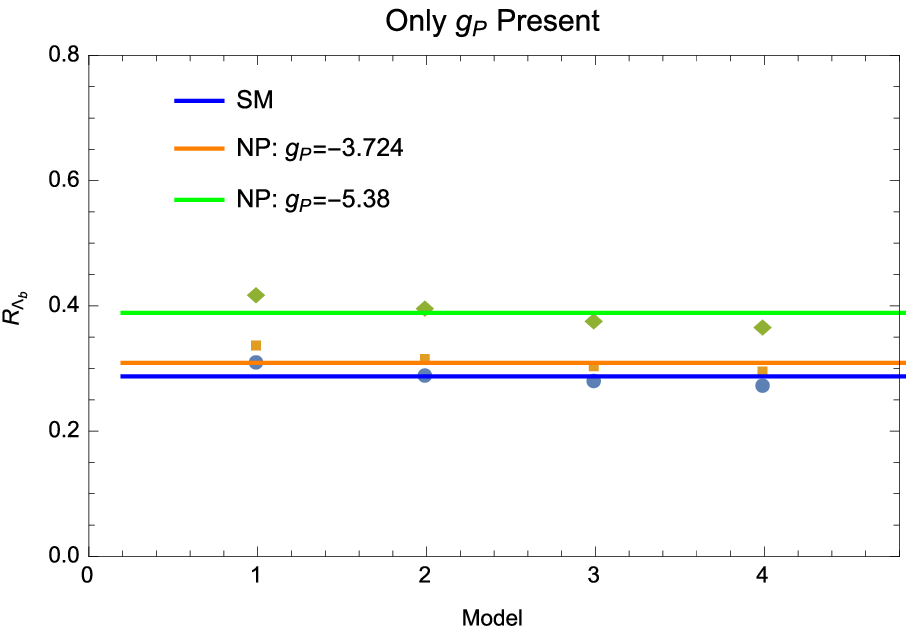}
\includegraphics[width=6.5cm, height=5.5cm]{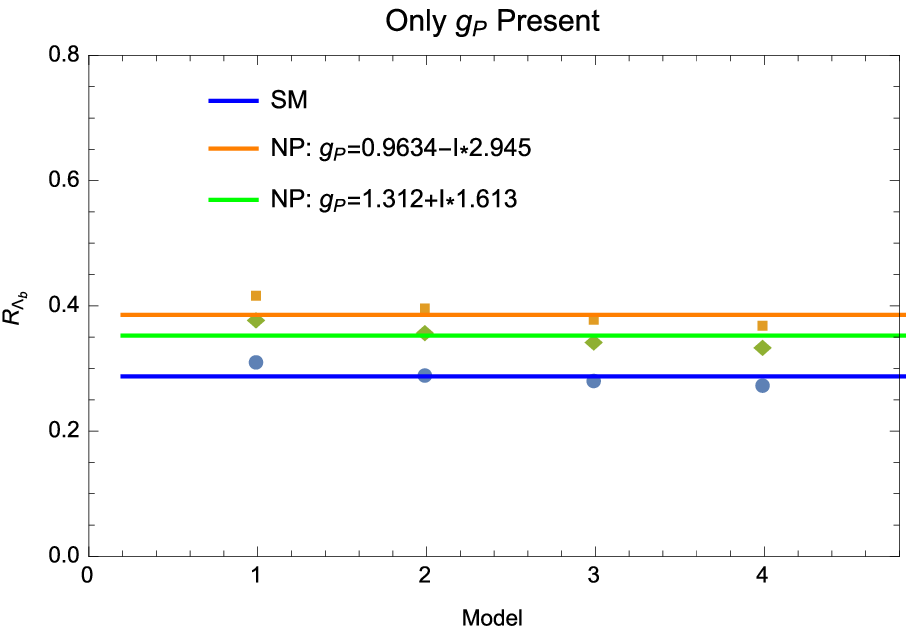}
\includegraphics[width=6.5cm, height=5.5cm]{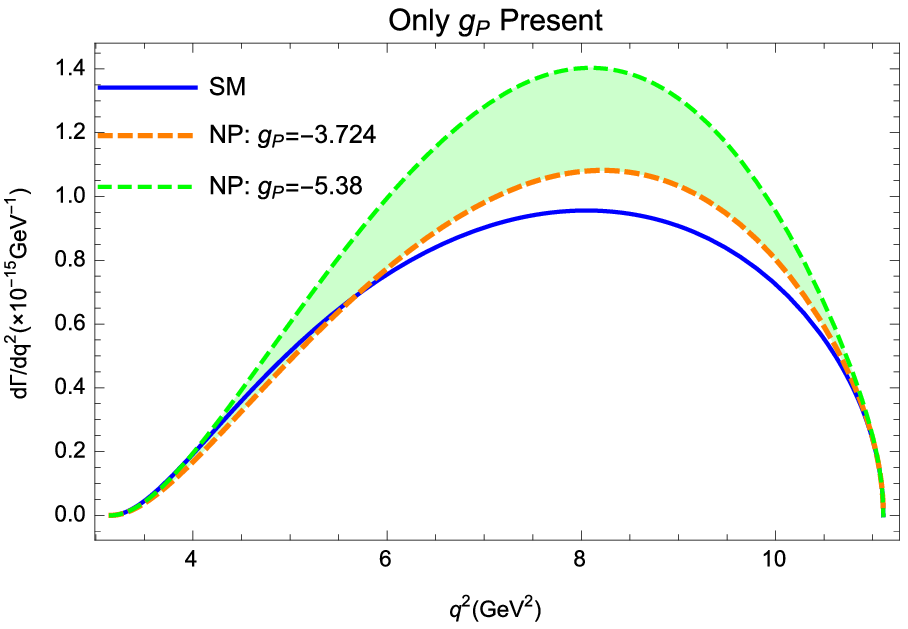}
\includegraphics[width=6.5cm, height=5.5cm]{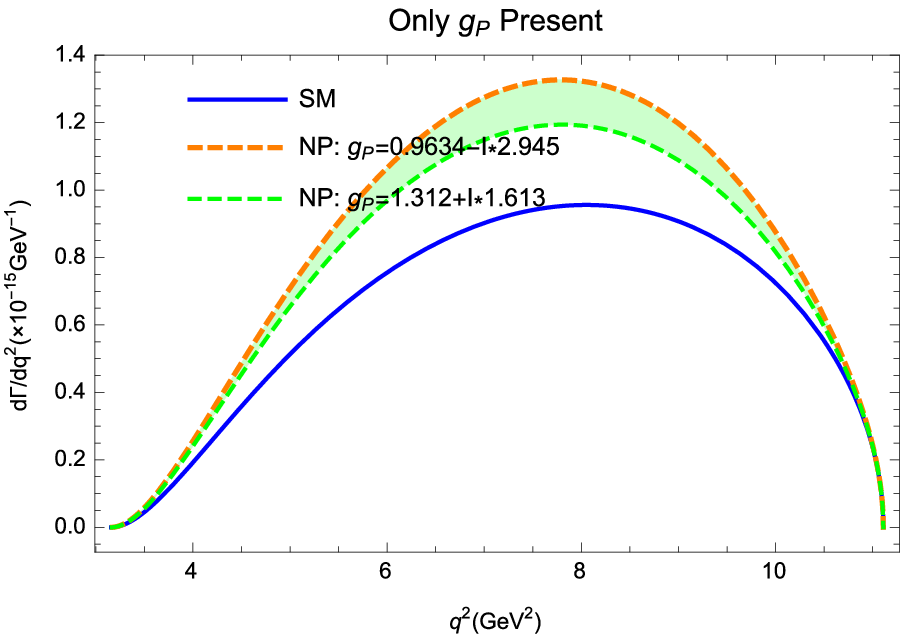}
\includegraphics[width=6.5cm, height=5.5cm]{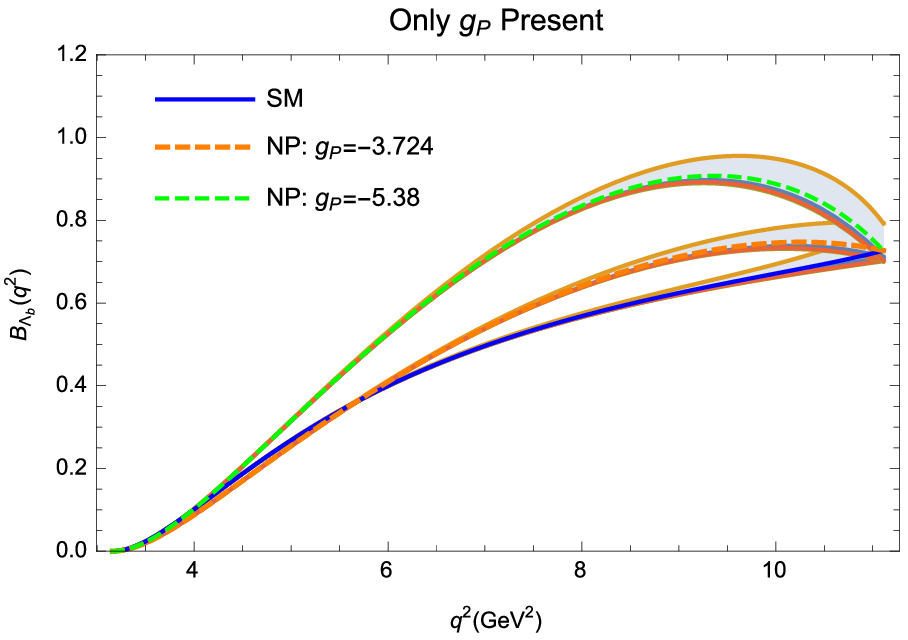}
\includegraphics[width=6.5cm, height=5.5cm]{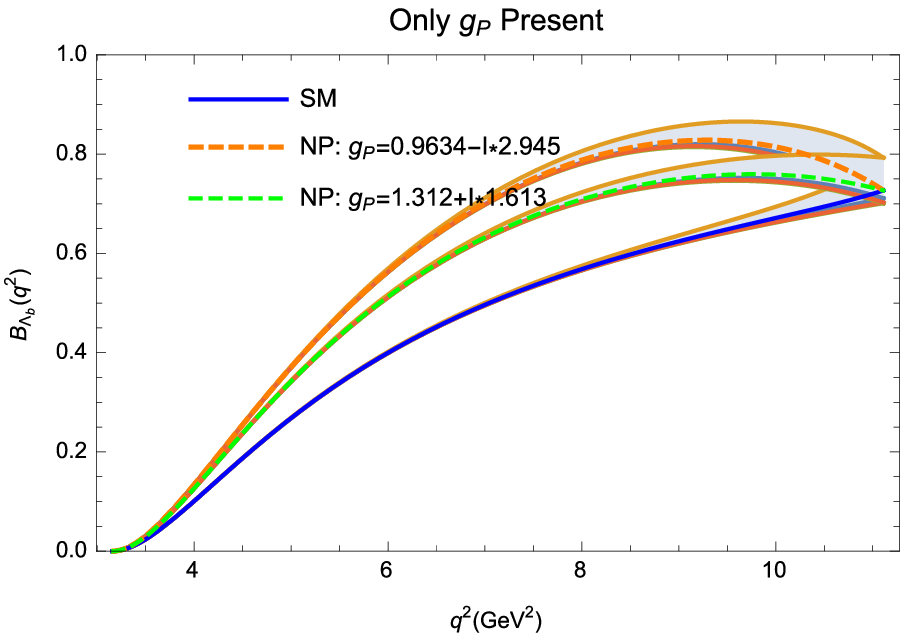}
\end{center}
\caption{The figures show the compared results between the standard model and new physics with only $g_{P}$ present.
The top and bottom row of graphs depict
$R_{\lb}=BR[{\Lambda_{b}\to\Lambda_{c}\tau\bar{\nu}_{\tau}}]/BR[{\Lambda_{b}\to\Lambda_{c}\ell\bar{\nu}_{\ell}}]$ 
and the ratio of differential distributions $B_{\lb}(q^2)=\frac{d\Gamma}{dq^{2}}(\Lambda_{b}\to\Lambda_{c}\tau\bar{\nu}_{\tau})/\frac{d\Gamma}{dq^{2}}(\Lambda_{b}\to\Lambda_{c}\ell\bar{\nu}_{\ell})$ as a function of $q^{2}$, respectively for the various form factors in Table \ref{FF}. The middle graphs depict 
the average differential decay rate with respect to $q^2$ for  the process $\Lambda_{b}\to\Lambda_{c}\tau\bar{\nu}_{\tau}$. Some representative values of the couplings have been chosen. 
}
\label{gP1}                                                                                                                                                                                                                                                                                                                                                                                                                                                                                                                                                                                                                                 
\end{figure}

Now I give the discussions of the results. From Eq.~(\ref{Efm}), we can make some general observations. We start with the case where only $g_{L}$ is present. In this case the NP Feynman amplitude has the same structure as the SM one and the total Feynman amplitude is just the SM amplitude modified by the factor $(1+g_L)$. Hence, if only $g_L$ is present, then
\bea
\label{gLratio}
R_{\lb} & = & {R_{\lb}}_{SM}|1+g_L|^2. \
\eea
Therefore, in this case, $R_{\lb} \ge  {R_{\lb}}_{SM}$ and we find the  range of $R_{\lb}$ to be $0.31 \backsim 0.44$. The shape of the differential distribution $d\Gamma/dq^2$ is the same as in the SM. In the left-side figures of Fig.~\ref{gL}, we show the plots for $R_{\lb}$, $d\Gamma/dq^2$ and $B_{\lb}(q^2)$ when only $g_L$ is present. We then consider the case where only $g_{R}$ is present. If only $g_R$ is present, then from Eq.~(\ref{Ehsm}) and Eq.~(\ref{EgR}) we can deduce that no clear relation between $R_{\lb}$ and  ${R_{\lb}}_{SM}$ can be obtained in this case. However, for the allowed $g_R$ couplings, we find $R_{\lb}$ is greater than the SM value and is in the range $0.30 \backsim 0.51$. The shape of the differential distribution $d\Gamma/dq^2$ is also the same as that of the SM. In the right-side figures of Fig.~\ref{gL}, we show the plots for $R_{\lb}$, $d\Gamma/dq^2$ and $B_{\lb}(q^2)$ when
only $g_R$ is present. 

We now move to the case when  only $g_{S,P}$ are present. Using Eq.~(\ref{Efm}), Eq.~(\ref{Efmsm}) and Eq.~(\ref{EgSP}), we can write
\bea
\label{SPratio}
R_{\lb} & = & {R_{\lb}}_{SM} + |g_S|^2 A_S + 2 Re(g_S) B_S,\nonumber\\
R_{\lb} & = & {R_{\lb}}_{SM} + |g_P|^2 A_P + 2 Re(g_P) B_P,\
\eea
since the physics of the the decay process only underlies in the Feyman amplitudes. The quantities $A_{S,P}$ and $B_{S,P}$ depend on masses and form factors and they are positive. Hence, for $ Re(g_P) \ge 0$ or  $ Re(g_S) \ge 0$, $R_{\lb}$ is always greater than or equal to  ${R_{\lb}}_{SM}$. But, for $ Re(g_P) < 0$ or  $ Re(g_S) < 0$, $R_{\lb}$ can be possibly less than the SM value. However, for the given constraints on $g_{S}$ here,  we can make $R_{\lb}$ only slightly less than the SM value while for $g_P$ it is always larger than the SM value. We find $R_{\lb}$ is in the range $0.28 \backsim 0.36$ when only $g_S$ is present and in the range $0.30 \backsim 0.42$ when only $g_P$ is present. In  Fig.~\ref{gS1} we show the plots for $R_{\lb}$, $d\Gamma/dq^2$ and $B_{\lb}(q^2)$ when only $g_S$ is present. The shape of the differential distribution $d\Gamma/dq^2$ can be different from that of the SM. In  Fig.~\ref{gP1} we show the plots for $R_{\lb}$, $d\Gamma/dq^2$ and $B_{\lb}(q^2)$ when only $g_P$ is present. In this case also the shape of the differential distribution $d\Gamma/dq^2$ can be different from that of the SM.

\begin{table}[tbh]
\center
\begin{tabular}{|c|c|c|}\hline
NP & $R_{\lb, min} $ & $R_{\lb, max} $ \\
\hline
Only $g_L$ & $0.31 $, $g_L=-0.065 +0.447 \ i $ & $0.44$, $g_L=-0.144 +0.903 \ i $\\
 Only $g_R$& $0.30 $, $g_R=-0.033 +0.119 \ i $ &   $0.51$, $g_R=0.182 +0.914 \ i$\\
Only $g_S$ & $0.28 $, $g_S=-1.442$ & $0.36$, $g_S=0.443$ \\
Only $g_P$ & $0.30 $, $g_P=0.587$ & $0.42$, $g_P=-5.859$ \\
\hline
\end{tabular}
\caption{Minimum and Maximum Values for the Averaged $R_{\lb}$.}
\label{Rminmax}
\end{table}

In Table~\ref{Rminmax}, we show the minimum and maximum values for the averaged $R_{\lb}$ with the corresponding NP couplings.

%%%%%%%%%%%%%%%%%%%%%%%%%%%%%%%%%%%%%%%%%%%%%%%%%%%%%%%%%%%%%%%%%%%%%%%
%%%%%%%%%%%%%%%%%%%%%%%%%%%%%%%%%%%%%%%%%%%%%%%%%%%%%%%%%%%%%%%%%%%%%%%%
\doublespacing
\chapter{CONCLUSION}

\singlespacing
\hspace*{\parindent}In this thesis, I calculated the SM and the NP predictions for the decay $\lblc$. Motivation to study this decay comes from the recent hints of lepton flavor non-universality observed by the BaBar Collaboration in $R(D^{(*)}) \equiv
  \frac{{\cal B}({\bar B} \to D^{(*)+} \tau^- {\bar\nu}_\tau)}{ {\cal
    B}({\bar B} \to D^{(*)+} \ell^- {\bar\nu}_\ell)}$ ($\ell =
  e,\mu$). I used a general parametrization of the NP operators and fixed the new physics couplings from the experimental measurements of $R(D)$ and $R(D^*)$. The predictions for $R_{\Lambda_b}$ (Eq.(\ref{Eratio1})), $\frac{d\Gamma}{dq^2}$, and $B_{\Lambda_b}(q^2)$ (Eq.(\ref{Eratio2})) are made by taking one of the various NP couplings at a time. We found the interesting results that $g_{L,R,P}$ couplings gave predictions larger than the SM values for all the three observables while the $g_{S}$ couplings gave  predictions which could be larger or smaller than the SM values. 

This thesis is related to our recent work of $\lblc$ decay  \cite{SWD}.

%%%%%%%%%%%%%%%%%%%%%%%%%%%%%%%%%%%%%%%%%%%%%%%%%%%%%%%%%%%%%%%%%%%%%%%%%
%appendix----------------------------------------------------------------

%%%%%%%%%%%%%%%%%%%%%%%%%%%%%%%%%%%%%
%reference--------------------------------------------------------------------------

\clearpage
\textbf{\addcontentsline{toc}{chapter}{BIBLIOGRAPHY}}
\begin{center}
\vspace*{\fill}
BIBLIOGRAPHY
\vspace*{\fill}
\end{center}

%%%%%%%%%%%%%%%%%%%%%%%%%%%%%%%%%%%%%%%%%%%%%%%%%%%%%%%%%%%%%%%%%%%%%%
%%%%%%%%%%%%%%%%%%%%%%%%%%%%%%%%%%%%%%%%%%%%%%%%%%%%%%%%%%%%%%%%%%%%%%%
%%%%%%%%%%%%%%%%%%%%%%%%%%%%%%%%%%%%%%%%%%%%%%%
%LIST OF APPENDICES

%\phantomsection
\clearpage
\textbf{\addcontentsline{toc}{chapter}{LIST OF APPENDICES}}
\begin{center}
\vspace*{\fill}
LIST OF APPENDICES
\vspace*{\fill}
\end{center}

\newpage
\doublespacing
\setlength\topmargin{-36 pt}
\begin{appendix}

%%%%%%%%%%%%%%%%%%%%%%%%%%%%%%%%%%%%%%%%%%%%%%%%%
%%											
%%  				APPENDIX A						
%%
%%%%%%%%%%%%%%%%%%%%%%%%%%%%%%%%%%%%%%%%%%%%%%%%%%%%%
%
\setlength{\oddsidemargin}{0in}
\setlength{\textwidth}{6.5in}
\setlength{\topmargin}{0in}
\setlength{\textheight}{9in}

\titlespacing*{\chapter}{0pt}{285pt}{12pt}% NEW
\titleformat{\chapter}[hang]% NEW
    {\normalsize\centering}{\MakeUppercase{\chaptertitlename}\ \thechapter:~}{0pt}{\normalsize}% NEW
\newpage

\chapter{NP OPERATORS EXPRESSED IN TERMS OF FORM FACTORS}
\label{npo}
\newpage
\setlength{\oddsidemargin}{0in}
\setlength{\textwidth}{6.5in}
\setlength{\topmargin}{0in}
\setlength{\textheight}{9in}

\singlespacing
If we consider the hadronic current:
\begin{equation}
\langle\Lambda_{c}|\bar{c}\gamma_{\mu}b|\Lambda_{b}\rangle=\bar{u}_{\lambda_{c}}(f_{1}\gamma_{\mu}+i f_{2}\sigma_{\mu\nu}q^{\nu}+f_{3}q_{\mu})u_{\lambda_{b}},
\end{equation}
then ($q=p-p_{3}$)
\begin{equation}
q^{\mu}\langle\Lambda_{c}|\bar{c}\gamma_{\mu}b|\Lambda_{b}\rangle=q^{\mu}\bar{u}_{\lambda_{c}}(f_{1}\gamma_{\mu}+i f_{2}\sigma_{\mu\nu}q^{\nu}+f_{3}q_{\mu})u_{\lambda_{b}}.
\label{hc1}
\end{equation}

The Left-Hand-Side of Eq. (\ref{hc1}) is
\begin{eqnarray}
q^{\mu}\langle\Lambda_{c}|\bar{c}\gamma_{\mu}b|\Lambda_{b}\rangle&=&\langle\Lambda_{c}|\bar{c}q^{\mu}\gamma_{\mu}b|\Lambda_{b}\rangle\nonumber\\&=&\langle\Lambda_{c}|\bar{c}\slashed{q}b|\Lambda_{b}\rangle \nonumber\\&=&\langle\Lambda_{c}|\bar{c}(\slashed{p}-\slashed{p_{3}})b|\Lambda_{b}\rangle \nonumber\\&=&(m_{b}-m_{c})\langle\Lambda_{c}|\bar{c}b|\Lambda_{b}\rangle,
\end{eqnarray}
where I used the equation of motion: $\slashed{p}b=m_{b}b$ and $\bar{c}\slashed{p_{3}}=m_{c}\bar{c}$, while the Right-Hand-Side of Eq.~(\ref{hc1}) is
\begin{eqnarray}
q^{\mu}\bar{u}_{\lambda_{c}}(f_{1}\gamma_{\mu}+i f_{2}\sigma_{\mu\nu}q^{\nu}+f_{3}q_{\mu})u_{\lambda_{b}}&=&\bar{u}_{\lambda_{c}}(f_{1}\slashed{q}+0+f_{3}q^2)u_{\lambda_{b}}.
\end{eqnarray}
Thus, we can get:
\begin{equation}
\langle\Lambda_{c}|\bar{c}b|\Lambda_{b}\rangle=\bar{u}_{\lambda_{c}}(f_{1}\frac{\slashed{q}}{m_{b}-m_{c}}+f_{3}\frac{q^2}{m_{b}-m_{c}})u_{\lambda_{b}}.
\end{equation}

Now, consider 
\begin{equation}
\langle\Lambda_{c}|\bar{c}\gamma_{\mu}\gamma_{5}b|\Lambda_{b}\rangle=\bar{u}_{\lambda_{c}}(g_{1}\gamma_{\mu}\gamma_{5}+i g_{2}\sigma_{\mu\nu}q^{\nu}\gamma_{5}+g_{3}q_{\mu}\gamma_{5})u_{\lambda_{b}},
\end{equation}
then
\begin{equation}
q^{\mu}\langle\Lambda_{c}|\bar{c}\gamma_{\mu}\gamma_{5}b|\Lambda_{b}\rangle=q^{\mu}\bar{u}_{\lambda_{c}}(g_{1}\gamma_{\mu}\gamma_{5}+i g_{2}\sigma_{\mu\nu}q^{\nu}\gamma_{5}+g_{3}q_{\mu}\gamma_{5})u_{\lambda_{b}}.
\label{hc2}
\end{equation}

The Left-Hand-Side of Eq. (\ref{hc2}) is
\begin{eqnarray}
q^{\mu}\langle\Lambda_{c}|\bar{c}\gamma_{\mu}\gamma_{5}b|\Lambda_{b}\rangle&=&\langle\Lambda_{c}|\bar{c}q^{\mu}\gamma_{\mu}\gamma_{5}b|\Lambda_{b}\rangle \nonumber\\&=&\langle\Lambda_{c}|\bar{c}\slashed{q}\gamma_{5}b|\Lambda_{b}\rangle\nonumber \\&=&\langle\Lambda_{c}|\bar{c}(\slashed{p}-\slashed{p_{3}})\gamma_{5}b|\Lambda_{b}\rangle\nonumber\\&=&-(m_{b}+m_{c})\langle\Lambda_{c}|\bar{c}\gamma_{5}b|\Lambda_{b}\rangle,
\end{eqnarray}
Where I used the equation of motion: $\slashed{p}b=m_{b}b$ and  $\bar{c}\slashed{p_{3}}=m_{c}\bar{c}$, and $\slashed{p}\gamma_{5}=-\gamma_{5}\slashed{p}$. The Right-Hand-Side of Eq.~(\ref{hc2}) is
\begin{eqnarray}
q^{\mu}\bar{u}_{\lambda_{c}}(g_{1}\gamma_{\mu}\gamma_{5}+i g_{2}\sigma_{\mu\nu}q^{\nu}\gamma_{5}+g_{3}q_{\mu}\gamma_{5})u_{\lambda_{b}}=\bar{u}_{\lambda_{c}}(g_{1}\slashed{q}\gamma_{5}+0+g_{3}q^{2}\gamma_{5})u_{\lambda_{b}}.
\end{eqnarray}
Thus, we can get:
\begin{equation}
\langle\Lambda_{c}|\bar{c}\gamma_{5}b|\Lambda_{b}\rangle=\bar{u}_{\lambda_{c}}(-g_{1}\frac{\slashed{q}\gamma_{5}}{m_{b}+m_{c}}-g_{3}\frac{q^2\gamma_{5}}{m_{b}+m_{c}})u_{\lambda_{b}}.
\end{equation}

\chapter{KINEMATICS}
\label{Akinematics}

\newpage

In the rest frame of $\Lambda_{b}$, we have:
\bea
p&=&(m,0,0,0),\nn\\
p_{1}&=&(E_{1},\vec{p}_{1}),\nn\\
p_{2}&=&(E_{2},\vec{p}_{2}),\nn\\
p_{3}&=&(E_{3},\vec{p}_{3}).
\eea

The transferred momentum $q=p_{1}+p_{2}=p-p_{3}$. We have $p_{3}=p-p_{1}-p_{2}$. By considering the Lorentz invariance, we can find out the following kinematic relations:
\bea
p^2&=&m^2,\nn\\
p_{1}^{2}&=&m_{1}^2,\nn\\
p_{2}^{2}&=&0,\nn\\
p_{3}^{2}&=&m_{3}^2,\nn\\
p\cdot p_{1}&=&m E_{1},\nn\\
p\cdot p_{2}&=&m E_{2},\nn\\
p\cdot p_{3}&=&m E_{3},\nn\\
p\cdot q&=&m E_{1}+m E_{2},\nn\\
p_{1}\cdot q&=&\frac{1}{2}(q^{2}+m_{1}^2),\nn\\
p_{1}\cdot p_{2}&=&\frac{1}{2}(q^{2}-m_{1}^2),\nn\\
p_{1}\cdot p_{3}&=&m E_{1}-\frac{1}{2}m_{1}^2-\frac{1}{2}q^2,\nn\\
p_{2}\cdot q&=&\frac{1}{2}(q^{2}-m_{1}^2),\nn\\
p_{2}\cdot p_{3}&=&m E_{2}+\frac{1}{2}m_{1}^2-\frac{1}{2}q^2,\nn\\
p_{3}\cdot q&=&\frac{1}{2}(q^{2}-m_{3}^2).
\eea

To achieve the integration of differential decay rate, let's define $p_{ij}=p_{i}+p_{j}$ and $m_{ij}^{2}=p_{ij}^2$. Then $m_{12}^{2}+m_{23}^{2}+m_{13}^{2}=m^{2}+m_{1}^{2}+m_{2}^{2}+m_{3}^{2}$ and $m_{12}^{2}=(p-p_{3})^2=m^{2}+m_{3}^{2}-2mE_{3}$, where $E_{3}$ is the energy of particle 3 in the rest frame of $m$.

From $m_{23}^{2}=(p-p_{1})^2=m^{2}+m_{1}^{2}-2mE_{1}$, we have
\bea
E_{1}&=&\frac{m^{2}+m_{1}^{2}-m_{23}^{2}}{2m}.
\eea

From $m_{13}^{2}=(p-p_{2})^2=m^{2}+m_{2}^{2}-2mE_{2}$ and $m_{12}^{2}+m_{23}^{2}+m_{13}^{2}=m^{2}+m_{1}^{2}+m_{2}^{2}+m_{3}^{2}$, we have
\bea
E_{2}&=&\frac{m_{12}^{2}+m_{23}^{2}-m_{1}^{2}-m_{3}^{2}}{2m}.
\eea

From $m_{12}^{2}=(p-p_{3})^2=m^{2}+m_{3}^{2}-2mE_{3}$, we have
\bea
E_{3}&=&\frac{m^{2}+m_{3}^{2}-m_{12}^{2}}{2m}.
\eea
Using the standard form for the Dalitz plot, we can get
\begin{eqnarray}
d\Gamma&=&\frac{(2\pi)^{4}}{2m}\bar{\lvert{M}\rvert^{2}}d\Phi_{n}(p;p_{1},...,p_{n})\nn\\&=&\frac{1}{(2\pi)^{3}}\frac{1}{8m}\bar{\lvert{M}\rvert^{2}}dE_{1}dE_{2}\nn\\&=&\frac{1}{(2\pi)^{3}}\frac{1}{32m^{3}}\bar{\lvert{M}\rvert^{2}}dm_{12}^{2}dm_{23}^{2}.
\end{eqnarray}
Here, the Dalitz plot:
for a given value of $m_{12}^{2}$, the range of $m_{23}^{2}$ is determined by its values when $\vec{p}_{2}$ is parallel or anti-parallel to $\vec{p}_{3}$ :
\bea
(m_{23}^{2})_{max}&=&(E_{2}^{*}+E_{3}^{*})^{2}-(\sqrt{{E_{2}^{*}}^2-m_{2}^{2}}-\sqrt{{E_{3}^{*}}^2-m_{3}^{2}})^{2},\nn\\
(m_{23}^{2})_{min}&=&(E_{2}^{*}+E_{3}^{*})^{2}-(\sqrt{{E_{2}^{*}}^2-m_{2}^{2}}+\sqrt{{E_{3}^{*}}^2-m_{3}^{2}})^{2},
\eea
where $E_{2}^{*}=(m_{12}^{2}-m_{1}^{2}+m_{2}^{2})/2m_{12}$ and $E_{3}^{*}=(m^{2}-m_{12}^{2}-m_{3}^{2})/2m_{12}$ are the energies of particles 2 and 3 in the $m_{12}$ rest frame. Since $m_{12}^{2}=q^2$, the differential decay rate with respect to $dq^{2}$ is:
\bea
\frac{d\Gamma}{dq^{2}}&=&\frac{1}{(2\pi)^{3}}\frac{1}{32m^{3}}\bar{\lvert{M}\rvert^{2}}dm_{23}^{2}.
\eea

\chapter{$\bar{B}\to D^{*}$ FORM FACTORS}
\label{Abdsdecay}

\newpage

\section{$\bar{B}\to D^{*}\tau^{-}\bar{\nu}_{\mu}$ Angular Distribution}

\hspace*{\parindent}The  full $\bar{B}\to D^{*}\tau^{-}\bar{\nu}_{\mu}$ angular distribution is given by \cite{bdnew3}
\bea
\frac{d \Gamma^{D^*}}{dq^2 d\cos{\theta_l}}&=& N |p_{D^*}| \Big[ 2 |{\cal{H}}_0|^2  \sin^2{\theta_l} + (|{\cal{H}}_\parallel|^2+ |{\cal{H}}_\perp|^2)  (1 + \cos{\theta_l})^2 -4 Re[{\cal{A}}_\parallel {\cal{H}}^*_\perp] \cos{\theta_l} \nn\\ && + \frac{m^2_\tau}{q^2} \Big( 2 |{\cal{H}}_0 \cos{\theta_l}-{\cal{H}}_{tP}|^2  + (|{\cal{H}}_\parallel|^2+ |{\cal{H}}_\perp|^2)  \sin^2{\theta_l} \Big) \Big],
\label{Ebdangular}
\eea
where $\theta_l$ is the angle between the $D^*$ meson and the $\tau$ lepton three-momenta in the $q^2$ rest frame, $N = \frac{G^2_F |V_{cb}|^2 q^2}{256 \pi^3 m^2_B} \Big(1-\frac{m_l^2}{q^2}\Big)^2$ and the amplitude ${\cal{H}}_{0,\parallel,\perp,t,P}$ are given in Sec. \ref{Samplitudes}. Also, the definition of ${\cal{H}}_{tP}$ is
\bea
{\cal{H}}_{tP} &=& \Big({\cal{H}}_t + \frac{\sqrt{q^2}}{m_\tau} {\cal{H}}_P \Big).
\eea

\section{$\bar{B}\to D^{*}\tau^{-}\bar{\nu}_{\mu}$ Form Factors}

\hspace*{\parindent}The relevant form factors for the $B \to D^*$ matrix elements of the vector $V_\mu = \bar{c}\gamma^{\mu}b$ and  axial-vector  $A_\mu = \bar{c}\gamma^{\mu} \gamma_5 b$ currents are defined as \cite{Rff}
\bea
\langle D^*|V_\mu |\bar{B}\rangle  &=&
 \frac{2 i V(q^2)}{m_B + m_{D^*}}\varepsilon_{\mu \nu \rho \sigma} \epsilon^{*\nu}  p^{\rho}_{D^*} p^{\sigma}_B \,,\nn\\
\langle D^*|A_\mu |\bar{B}\rangle&=&  2 m_{D^*} A_0 (q^2)\frac{\epsilon^*\cdot q}{q^2} q_\mu + (m_B + m_{D^*}) A_1(q^2) \Big[\epsilon^*_{\mu}-\frac{\epsilon^*\cdot q}{q^2} q_\mu \Big]\nn\\&& -A_2(q^2) \frac{\epsilon^*\cdot q}{(m_B + m_{D^*})} \Big[(p_B +p_{D^*})_\mu -\frac{m^2_B-m^2_{D^*}}{q^2}q_\mu \Big]\,.
\label{Evacurrent}
\eea

In the Heavy Quark Effective Theory (HQET), the form factors in Eq.~(\ref{Evacurrent}) are given by \cite{BDSM1, Rff2, Rff3}
\bea
A_0(q^2)& =&\frac{R_0(w)}{R_{D^*}}h_{A_1}(w), \nn\\
A_1(q^2) &= & R_{D^*} \frac{w+1}{2}h_{A_1}(w),\nn\\
A_2(q^2) &= & \frac{R_2(w)}{R_{D^*}}h_{A_1}(w),\nn\\
V(q^2) &=&\frac{R_1(w)}{R_{D^*}}h_{A_1}(w),
\eea
where $R_{D^*} = 2 \sqrt{m_B m_D^*}/(m_B + m_D^*)$. The summary results of $w$ dependence of the form factors \cite{BDSM1, Rff2} are
\bea
h_{A_1}(w) &=& h_{A_1}(1)\Big[1-8 \rho^2 z + (53 \rho^2-15)z^2  -(231 \rho^2 -91) z^3\Big],\nn\\
R_1(w)  &=& R_1(1) - 0.12 (w-1) + 0.05 (w-1)^2, \nn\\
R_2(w)  &=& R_2(1) + 0.11 (w-1) - 0.06(w-1)^2, \nn\\
R_0(w)  &=& R_0(1) - 0.11 (w-1) + 0.01(w-1)^2 ,
\eea
where $z =( \sqrt{w+1}-\sqrt{2})/( \sqrt{w+1}+\sqrt{2})$. The numerical values of the free parameters $\rho^2$, $h_{A_1}(1)$, $R_1(1)$ and $R_2(1) $ are \cite{Rff3}
\bea
h_{A_1}(1) |V_{cb}| &=& (34.6\pm  0.2 \pm 1.0) \times 10^{-3},\nn\\
\rho^2 &=& 1.214 \pm 0.034 \pm 0.009,\nn\\
R_1(1) &=& 1.401\pm 0.034 \pm 0.018,\nn\\
R_2(1) &=& 0.864 \pm 0.024 \pm 0.008,
\eea
and $R_0(1) = 1.14$ is taken from Ref.~\cite{BDSM1}. In the numerical analysis, we may allow  $10 \%$ uncertainties in the $R_0(1)$ value to account  higher order corrections.

Therefore, in the HQET the amplitudes in Eq.~(\ref{Ebdangular}) become
\bea
{\cal{H}}_{tP}&=&-m_B(1+r_*)\sqrt{\frac{r_* (w^2-1)}{1+r_*^2-2r_* w}}h_{A_1}(w)R_{0}(w)\big[ (1-g_A)+\frac{m_B^2(1+r_*^2-2r_* w)}{m_l(m_b+m_c)}g_P \big],\nn\\
{\cal{H}}_0  &=& -\frac{ m_B  (1 - r_*) (w + 1) \sqrt{r_*}}{ \sqrt{ (1 + r^2_* - 2 r_* w)}}h_{A_1}(w) \Big[1 + \frac{(w-1)(1-R_2(w))}{(1 - r_*)}\Big](1-g_A),\nn\\
{\cal{H}}_\parallel &=&  m_B \sqrt{2r_*}(w + 1)h_{A_1}(w)(1 - g_A),\nn\\
{\cal{H}}_\perp &=& m_B   \sqrt{2r_* (w^2 - 1)} h_{A_1}(w)R_1(w) (1 + g_V),
\eea
where $r_* = m_{D^*}/m_B$.

%%%%%%%%%%%%%%%%%%%%%%%%%%%%%%%%%%%%%%%%%%%%%%%%%%%%%%%%%%%%%%%%%%%%%%%%%%
%kinematics-------------------------------------------------------------
\section{$\bar{B}\to D^{*}$ Kinematics}

\hspace*{\parindent}Here, for $a=(a_0, a_1, a_2, a_3)$ and $b=(b_0, b_1, b_2, b_3)$, we have  $a \cdot b= a_0 b_0-(a_1 b_1 +a_2 b_2 + a_3 b_3)$.

In the rest frame of $\bar{B}$:
\bea
p_B&=&(m_B,0,0,0),\nn \\
p_{D^*}&=&(E_{{D^*}},0,0,|p_{D^*}|),\nn\\
q&=&(q_0,0,0,-|p_{D^*}|),
\eea
where $q=p_B=p_{D^*}$ and
\bea
E_{D^{*}}&=&(m_{B}^{2}+m_{D^{*}}^{2}-q^2)/(2 m_{B}),\nn\\
|p_{D^*}|&=&\sqrt{(\frac{m_{B}^{2}+m_{D^*}^{2}-q^2}{2m_B})^2-m_{D^{*}}^{2}},\nn\\
q_{0}&=&(m_{B}^{2}-m_{D^*}^{2}+q^2)/(2m_B).
\eea

The polarization vectors of $D^*$ are given by
\bea
\epsilon_0&=&\frac{1}{m_{D^*}}(|p_{D^*}|,0,0,E_{D^*}),\nn\\
\epsilon_\pm&=&\mp\frac{1}{\sqrt{2}}(0,1,\pm i,0).
\eea

The polarization vector of virtual gauge boson are given by
\bea
\bar{\varepsilon}_0&=&\frac{1}{\sqrt{q^2}}(|p_{D^*}|,0,0,-q_{0}),\nn\\
\bar{\varepsilon}_\pm&=&\frac{1}{\sqrt{2}}(0,\pm 1,-i,0),\nn\\
\bar{\varepsilon}_t&=&\frac{1}{\sqrt{q^2}}(q_{0},0,0,-|p_{D^*}|).
\eea

To get the amplitudes, we have the following useful relations:
\bea
\bar{\varepsilon}_{0}^* \cdot q &=&0, \nn\\
\bar{\varepsilon}_{0}^* \cdot p_{D^*} &=& \frac{1}{\sqrt{q^2}}(|p_{D^*}|E_{D^*}+|p_{D^*}|q_0)=\frac{1}{\sqrt{q^2}} m_B |p_{D^*}|,\nn\\
\eea
\bea
\epsilon_0^* \cdot q &=&\frac{1}{m_{D^*}}(|p_{D^*}|E_{D^*}+|p_{D^*}|q_0)=\frac{m_B |p_{D^*}|}{m_{D^*}}, \nn\\
\epsilon_0^* \cdot \bar{\varepsilon}_0^* &=& \frac{1}{m_{D^*}\sqrt{q^2}}(|p_{D^*}|^2+E_{D^*}q_0),
\eea
\bea
\bar{\varepsilon}_{t}^*\cdot q &=&\sqrt{q^2}, \nn\\
\epsilon_t^* \cdot \bar{\varepsilon}_t^* &=& \frac{1}{m_{D^*}}(|p_{D^*}|E_{D^*}+|p_{D^*}|q_0)=\frac{1}{m_{D^*}} m_B |p_{D^*}|,
\eea
\bea
\bar{\varepsilon}_{+}^*\cdot \epsilon_+^*&=&\frac{1}{\sqrt{2}}(0,1,i,0)\cdot [-\frac{1}{\sqrt{2}}(0,1,-i,0)]=-\frac{1}{2}(0-1-1-0)=1,\nn\\
\bar{\varepsilon}_{-}\cdot \epsilon_-^*&=&\frac{1}{\sqrt{2}}(0,-1,i,0)\cdot \frac{1}{\sqrt{2}}(0,1,i,0)=\frac{1}{2}(0+1+1+0)=1,
\eea
\bea
\bar{\varepsilon}_{\pm}^*\cdot q &=&0.
\eea

\section{$\bar{B}\to D^{*}$ Amplitudes}
\label{Samplitudes}

\hspace*{\parindent}In Ref.~\cite{bdnew4}, Eq.~(A.6) (the V-part) used $\epsilon_{\mu\nu\rho\sigma}$. I used $\epsilon_{0123}=1$, which I think should agree with in Ref.~\cite{Rff} where the authors used $\epsilon^{0123}=-1$.

\begin{enumerate}
\item \textbf{${\cal{H}}_0$:}
\bea
V_0&=&0,\nn\\
A_{0}&=&\bar{\varepsilon}_0^* \langle D^*|A_\mu|\bar{B}  \rangle \nn\\&=&(m_B+m_{D^*})A_{1}(q^2)\epsilon_0^* \cdot \bar{\varepsilon}_0^*-\frac{\epsilon_0^* \cdot q}{m_B+m_{D^*}}A_{2}(q^2)](2p_{D^*}\cdot \bar{\varepsilon}_0^*)\nn\\&=&-\frac{1}{2m_{D^*}\sqrt{q^2}}[(m_{B}^{2}-m_{D^*}^{2}-q^2)(m_B+m_{D^*})A_{1}(q^2)-\frac{4m_{B}^{2}|p_{D^*}|^2}{m_B+m_{D^*}}A_{2}(q^2)],
\eea
\bea
{\cal{H}}_0&=&(V_0-A_0) (1-g_A)\nn\\
&=&-\frac{1}{2m_{D^*}\sqrt{q^2}}[(m_{B}^{2}-m_{D^*}^{2}-q^2)(m_B+m_{D^*})A_{1}(q^2)\nn\\&&
-\frac{4m_{B}^{2}|p_{D^*}|^2}{m_B+m_{D^*}}A_{2}(q^2)](1-g_A).
\eea
Here, $|p_{D^*}|^2+q_0 E_{D^*}=\frac{m_{B}^{2}-m_{D^*}^{2}-q^2}{2}$.

\item \textbf{${\cal{H}}_t$:}
\bea
V_t&=&0,\nn\\
A_{t}&=&\bar{\varepsilon}_t^* \langle D^*|A_\mu|\bar{B}  \rangle \nn\\&=&\frac{2m_{D^*}A_0 (q^2)\epsilon_t^* \cdot q}{q^2}\sqrt{q^2}+(m_B+m_{D^*})A_{1}(q^2)(\epsilon_t^* \cdot \bar{\varepsilon}_t-\frac{\epsilon_t^* \cdot q}{\sqrt{q^2}})\nn\\&&-\frac{A_2 (q^2)\epsilon_t^* \cdot q}{m_B+m_{D^*}}(\frac{m_B^2-m_{D^*}^2}{\sqrt{q^2}}-\frac{m_B^2-m_{D^*}^2}{\sqrt{q^2}})\nn\\&=&\frac{2}{\sqrt{q^2}}m_{B}|p_{D^*}|A_{0}(q^2),
\eea
\bea
{\cal{H}}_t &=&(V_t -A_t)(1-g_A)= -\frac{2}{\sqrt{q^2}}m_{B}|p_{D^*}|A_{0}(q^2)(1-g_A).
\eea

\item \textbf{$H_{\pm}$:}
\bea
V_+ &=&\bar{\varepsilon}_+^* \langle D^*|V_\mu|\bar{B} \rangle\nn\\
&=&\frac{2iV(q^2)}{m_B+m_{D^*}}\varepsilon_{\mu\nu\rho\delta}\bar{\varepsilon}_+^{*\mu}{\epsilon^{*}}^{\nu}p_{D^*}^{\rho}p_B^{\delta}\nn\\
&=&\frac{2iV(q^2)}{m_B+m_{D^*}}(\varepsilon_{1230}\bar{\varepsilon}_+^{*1}{\epsilon^{*}}^{2}+\varepsilon_{2130}\bar{\varepsilon}_+^{*2}{\epsilon^{*}}^{1})p_{D^*}^{3}p_B^{0}\nn\\
&=&\frac{2iV(q^2)}{m_B+m_{D^*}}[-\frac{1}{\sqrt{2}}(\frac{i}{\sqrt{2}})+\frac{i}{\sqrt{2}}(\frac{-1}{\sqrt{2}})] |p_{D^*}|m_B\nn\\
&=&\frac{2V(q^2)m_{B}|p_{D^*}|}{m_B+m_{D^*}},\nn\\
A_+ &=&\bar{\varepsilon}_+^* \langle D^*|A_\mu|\bar{B} \rangle\nn\\
&=&(m_B+m_{D^*})A_{1}(q^2)\epsilon_+^* \cdot \bar{\varepsilon}_+^* -\frac{A_2 (q^2)\epsilon^* \cdot q}{m_B+m_{D^*}}* 0\nn\\
&=&(m_B+m_{D^*})A_1(q^2),
\eea
\bea
{\cal{H}}_+&=&V_+ (1+g_V) -A_+ (1-g_A)\nn\\
&=&\frac{2V(q^2)m_{B}|p_{D^*}|}{m_B+m_{D^*}}(1+g_V)-(m_B+m_{D^*})A_1(q^2)(1-g_A).
\eea
\bea
V_- &=&\bar{\varepsilon}_-^* \langle D^*|V_\mu|\bar{B}\rangle\nn\\
&=&\frac{2iV(q^2)}{m_B+m_{D^*}}\varepsilon_{\mu\nu\rho\delta}\bar{\varepsilon}_-^{^*\mu}{\epsilon^{*}}^{\nu}p_{D^*}^{\rho}p_B^{\delta}\nn\\
&=&\frac{2iV(q^2)}{m_B+m_{D^*}}(\varepsilon_{1230}\bar{\varepsilon}_-^{*1}{\epsilon^{*}}^{2}+\varepsilon_{2130}\bar{\varepsilon}_-^{*2}{\epsilon^{*}}^{1})p_{D^*}^{3}p_B^{0}\nn\\
&=&\frac{2iV(q^2)}{m_B+m_{D^*}}[\frac{1}{\sqrt{2}}(\frac{i}{\sqrt{2}})+\frac{i}{\sqrt{2}}(\frac{1}{\sqrt{2}})] |p_{D^*}|m_B\nn\\
&=&-\frac{2V(q^2)m_{B}|p_{D^*}|}{m_B+m_{D^*}},\nn\\
A_- &=&\bar{\varepsilon}_-^* \langle D^*|A_\mu|\bar{B} \rangle\nn\\
&=&(m_B+m_{D^*})A_{1}(q^2)\epsilon_-^* \cdot \bar{\varepsilon}_-^* -\frac{A_2 (q^2)\epsilon^* \cdot q}{m_B+m_{D^*}}* 0\nn\\
&=&(m_B+m_{D^*})A_1(q^2),
\eea
\bea
{\cal{H}}_-&=&V_-(1+g_V) -A_-(1-g_A)\nn\\
&=&-\frac{2V(q^2)m_{B}|p_{D^*}|}{m_B+m_{D^*}}(1+g_V)-(m_B+m_{D^*})A_1(q^2)(1-g_A).
\eea
Here, $\varepsilon_{1230}=-1$, $\varepsilon_{2130}=1$.

\item \textbf{${\cal{H}}_{(\parallel, \perp)}$:}
\bea
{\cal{H}}_{\perp}&=&\frac{1}{\sqrt{2}}({\cal{H}}_{+} - {\cal{H}}_{-})\nn\\
&=&2\sqrt{2}\frac{V(q^2)m_{B}|p_{D^*}|}{m_B+m_{D^*}}(1+g_V),
\eea
\bea
{\cal{H}}_{\parallel}&=&\frac{1}{\sqrt{2}}({\cal{H}}_+ + {\cal{H}}_-)\nn\\
&=&-\sqrt{2}(m_B+m_{D^*})A_1(q^2)(1-g_A).
\eea

\item \textbf{${\cal{H}}_{P}$:}
\bea
q^\mu \langle D^*|\bar{c}\gamma_{\mu} \gamma_5 b|\bar{B}\rangle&=&\langle D^*|\bar{c}\slashed{q} \gamma_5 b|\bar{B}\rangle\nn\\
&=&\langle D^*|\bar{c}(\slashed{p_B}-\slashed{p_{D^*}}) \gamma_5 b|\bar{B}\rangle\nn\\
&=&-(m_B+m_{D^*})\langle D^*|\bar{c}\gamma_5 b|\bar{B}\rangle.
\eea
Here, I already used the equation of motion. Also,
\bea
q^\mu \langle D^*|\bar{c}\gamma_{\mu} \gamma_5 b|\bar{B}\rangle &=&2m_{D^*}|p_{D^*}| A_{0}(q^2)\epsilon^* \cdot q\nn\\
&=&2m_{D^*}|p_{D^*}| A_{0}(q^2)\frac{m_B |p_{D^*}|}{m_{D^*}} ,
\eea
\bea
{\cal{H}}_P&=&\langle D^*|\bar{c}\gamma_5 b|\bar{B}\rangle g_P \nn\\
&=&\frac{-2m_B |p_{D^*}| A_{0}(q^2)}{m_B+m_{D^*}}g_P.
\eea

\end{enumerate}

%%%%%%%%%%%%%%%%%%%%%%%%%%%%%%%%%%%%%%%%%%%%%%%%%%%%%%%%%%%%%%%%%%%%%%%%

%%%%%%%%%%%%%%%%%%%%%%%%%%%%%%%%%%%%%%%%%%%%%%%%%%%%%%%%%%%%%%%%%%%%%%%%

%%%%%%%%%%%%%%%%%%%%%%%%%%%%%%%%%%%%%%%%%%%%%%%%%%%%%%%%%%%%%%%%%%%%%%%%
%BtoD-----------------------------------------------------------------

\chapter{$\bar{B}\to D\tau\bar{\nu}_\tau$ FORM FACTORS}
\label{Abddecay}

\newpage
\section{$\bar{B}\to D\tau\bar{\nu}_\tau$ Angular Distribution}

\hspace*{\parindent}The $\bar{B}\to D\tau\bar{\nu}_\tau$ angular distribution (the differential decay rate) for the lepton helicity $\lambda_\tau = \pm \frac{1}{2}$ are
\bea
\frac{d \Gamma^D[\lambda_\tau = -1/2]}{dq^2 d\cos{\theta_l}}&=& 2 N |p_D|  |H_0|^2  \sin^2{\theta_l}, \nn\\
\frac{d \Gamma^D[\lambda_\tau = 1/2]}{dq^2 d\cos{\theta_l}}&=& 2 N |p_D|  \frac{m^2_\tau}{q^2} |H_0 \cos{\theta_l}-H_{tS}|^2.
\eea

The differential decay rate corresponding to the helicity $\lambda_\tau = 1/2$  vanishes for the light leptons $(e,\mu)$ as their mass is much smaller.

\section{$\bar{B}\to D\tau\bar{\nu}_\tau$ Amplitudes}

\hspace*{\parindent}The methods to get the amplitudes from the  $B \to D$ matrix elements is the same as for  $B \to D^*$. Here, I didn't include the details.

The amplitudes are 
\bea
H_0 &=& \frac{2 m_B |p_D|}{\sqrt{q^2}} F_+(q^2) (1 + g_V ),\nn\\
H_t & =& \frac{m^2_B -m^2_D}{\sqrt{q^2}}  F_0(q^2)   (1 + g_V ),\nn\\
H_S &=& \frac{m^2_B -m^2_D}{m_b - m_c}  F_0(q^2)  g_S.
\label{Ebdam}
\eea
Also, the definition of $H_{tS}$ is 
\bea
H_{tS}&=&H_t+\frac{\sqrt{q^2}}{m_\tau}H_S.
\eea

\section{$\bar{B}\to D\tau\bar{\nu}_\tau$ Form Factors}

\hspace*{\parindent}The form factors $F_+(q^2)$ and $F_0(q^2)$ of the  $B \to D$ matrix elements are defined as
\bea
\langle D|\bar{c}\gamma^{\mu}b|\bar{B}\rangle &=&F_+(q^2)\Big[p_B^{\mu}+p_D^{\mu}-\frac{m_B^2-m^2_D}{q^2}q^{\mu}\Big] 
  +  F_0(q^2)~\frac{m_B^2-m^2_D}{q^2}q^{\mu},\nn\\
\langle D|\bar{c}b|\bar{B}\rangle  &=&  \frac{m_B^2-m^2_D}{
   m_b (\mu) - m_c(\mu)}F_0(q^2).
\label{Ebdff}
\eea
In the heavy quark effective theory, the form factors in Eq.(\ref{Ebdff}) are given by
\bea
F_+(q^2) &=& \frac{V_1(w)}{R_D},\nn\\
F_0(q^2) &=& \frac{(1 + w) R_D}{2}  S_1(w),
\eea
where $R_{D} = 2 \sqrt{m_B m_D}/(m_B + m_D)$ and $r = m_D/m_B$.
The parametrization of the  form factor $V_1(w)$  is given by \cite{Rff2}
\bea
V_1(w)  &=& V_1(1) [1 -8 \rho^2_1 z + (51 \rho^2_1 - 10)z^2  - (252 \rho^2_1- 84) z^3 ],
\eea
where $z =( \sqrt{w+1}-\sqrt{2})/( \sqrt{w+1}+\sqrt{2})$. The numerical values of the free parameters are \cite{Rff5}
\bea
V_1(1) |V_{cb}| &=& (43.0 \pm 1.9 \pm 1.4)\times 10^{-3},\nn\\
\rho^2_1 &=& 1.20 \pm 0.09 \pm 0.04.
\eea
The parametrization of form factor $ S_1(w)$ is given by \cite{BDSM2}
\bea
S_1(w)&=&1.0036 [1- 0.0068(w-1)+ 0.0017(w-1)^2-0.0013(w-1)^3] V_1(w).
\eea

In the HQET, the amplitudes in Eq. (\ref{Ebdam}) becomes
\bea
H_0 &=& m_B (1 + r) \sqrt{\frac{r (w^2 - 1)}{(1 + r^2 - 2 r w)}} V_1[w](1 + g_V),\nn\\
H_{tS} &=& \frac{m_B (1 - r) \sqrt{r} (w + 1)}{\sqrt{(1 + r^2 - 2 r w)}}    S_1[w]  \Big[(1 + g_V)+\frac{m_B^2 (1 + r^2 - 2 r w) }{m_l(m_b(\mu) - m_c(\mu)) } g_S \Big].
\eea

\clearpage
\end{appendix}

%\phantomsection
\newpage
\newgeometry{left=1in, right=1in, top=2in, bottom=1in}

\begin{center}
VITA
\textbf{\addcontentsline{toc}{chapter}{VITA}}
\end{center}
\singlespacing

\begin{center}
	\begin{flushleft}
	Wanwei Wu \hspace{89mm}Email: wwu1@go.olemiss.edu
	\end{flushleft}
	\rule{\textwidth}{0.5pt}
\end{center}

\vspace*{10pt}

\begin{flushleft}
EDUCATION
\begin{itemize}
\item{M.S. in Physics, University of Mississippi, Oxford, MS, August 2013-May 2015}\\
      Thesis: \textit{Semi-leptonic Decay of Lambda-b in the Standard Model and With New Physics}

\item B.S. in Physics, Sichuan University, Chengdu, China, September 2006-June 2010      
      
\end{itemize}
\end{flushleft}

\begin{flushleft}
TEACHING EXPERIENCE
	\begin{itemize}
	\item Teaching Assistant, August 2013-May 2015\\
	Department of Physics and Astronomy, University of Mississippi\\
	Courses: Phys223 (Physics Lab), Phys652 (Mathematical Physics)
	\end{itemize}

\end{flushleft}

\begin{flushleft}
SUMMER SCHOOL
	\begin{itemize}
	\item The General Theory of Relativity---Theory and Experiment Graduate Summer School\\
	Huazhong University of Science and Technology, Wuhan, China, June-July 2009
	\end{itemize}
	
\end{flushleft}

\begin{flushleft}
PROGRAMMING SKILLS
\begin{itemize}
\item Mathematica, MATLAB, Python, C/C++, Fortran
\end{itemize}
\end{flushleft}

\begin{flushleft}
INTEREST
\begin{itemize}
\item Literature, Chess, Travelling, Woodworking
\end{itemize}
\end{flushleft}

\clearpage

\end{document}